\documentclass[a4paper,12pt,twoside]{report}
\usepackage[left=3cm,right=3cm,top=3cm,bottom=3cm]{geometry} 
\usepackage{pdfpages}
\usepackage{hyperref}
\usepackage{listings}
\usepackage{xcolor}
\usepackage{setspace}
\usepackage{tocloft}
\usepackage{amsmath, amsfonts,amssymb}
\usepackage{chngcntr}
\usepackage[toc,page]{appendix}
\usepackage[T1]{fontenc}
\usepackage[nottoc]{tocbibind}
\usepackage[compact]{titlesec}
\usepackage{nicematrix}
\usepackage{MnSymbol}
\usepackage{multirow}
\usepackage{graphicx} 
\usepackage{graphics} 
\usepackage{subcaption}
\usepackage{booktabs} 

\titlespacing{\section}{0pt}{1ex}{0ex}
\titlespacing{\subsection}{0pt}{1ex}{0ex}
\titlespacing{\subsubsection}{0pt}{1ex}{0ex}

\counterwithout{figure}{chapter}
\counterwithout{table}{chapter}

\usepackage{algorithm}    
\usepackage{algpseudocode}
\usepackage{graphicx}
\usepackage{verbatim}
\usepackage{latexsym}
\input{preamble/mathchars.sty}
\usepackage{setspace}
\usepackage{blindtext}
\usepackage{float}
\usepackage{enumitem}
\usepackage{nomencl}
\usepackage{etoolbox}  

\input{preamble/blocked.sty}
\input{preamble/uhead.sty}
\input{preamble/boxit.sty}
\input{preamble/icthesis.sty}

\newcommand{\narrowlinespacing}{\renewcommand{\baselinestretch}{1.0} \normalsize}

\cftsetindents{section}{0in}{0.5in}
\cftsetindents{subsection}{0in}{0.5in}
\cftsetindents{subsubsection}{0in}{0.5in}
\cftsetindents{paragraph}{0in}{0.5in}

\makeatletter
\def\l@subsection{\@dottedtocline{2}{1.5em}{3.2em}}
\def\l@subsubsection{\@dottedtocline{3}{3.8em}{4.1em}}
\makeatother

\renewcommand\nomgroup[1]{%
  \item[\bfseries
  \ifstrequal{#1}{N}{Number sets}{%
  \ifstrequal{#1}{P}{Parameters}{%
  \ifstrequal{#1}{O}{Operators}{}}}%
]}

\renewcommand{\nompreamble}{%
  \begin{spacing}{1} 
}
\renewcommand{\nompostamble}{%
  \end{spacing}
}

\makenomenclature

\usepackage[toc,page]{appendix}

\narrowlinespacing

\setlist[itemize]{itemsep=0.5ex, topsep=0.5ex}
\setlist[enumerate]{itemsep=0.5ex, topsep=0.5ex}

\usepackage{natbib}
\let\cite\citep

\begin{document}

\newgeometry{left=2cm,right=2cm} 

\title{“Revisiting Meter Tracking in Carnatic Music using Deep Learning Approaches”}
\author{Satyajeet Prabhu}
\submitdate{August 2025}
\supervisor{Martín Rocamora}
\cosupervisor{Thomas Nuttall}

\maketitle
\restoregeometry

\cleardoublepage 

\chapter*{Acknowledgments}
I would like to express my sincere gratitude to Prof. Xavier Serra for giving me the opportunity to be part of this prestigious program despite my limited experience in software development. His encouragement to explore research in Indian Art Music has been a source of inspiration for me and many aspiring music computation researchers from India.

I am deeply grateful to my supervisor, Dr. Martín Rocamora, whose constant guidance over the two years of this program has been invaluable. From teaching one of the most engaging courses in the program to offering me an internship opportunity at the MTG and ultimately supervising my thesis, he has played a pivotal role in shaping me as a researcher.

I am thankful to my supervisor Thomas Nuttall - Tom, as he is affectionately known - whose support began even before the program, sparked by our meeting at the ISMIR satellite workshop in India in 2022. My gratitude also extends to Genís, Adithi, Oguz, Behzad, Esteban, Jyoti, Alia and all the other PhD students and researchers at MTG, who have always been willing to offer assistance and guidance in both professional and personal matters.

I also want to sincerely thank Ajay Srinivasamurthy, the author of the work on which this study is based, for being generous with his time and constantly offering his support despite his busy schedule.

It has been a privilege to study alongside my incredibly talented colleagues in the SMC Masters program, who I now proudly call friends. Special thanks to Anmol Mishra, now my co-author as well, for the banter and for the constant encouragement to take on new challenges and to Robin Doerfler for some of the most philosophical and intellectually stimulating conversations I have ever had.

Lastly, I am immensely thankful to my parents for their unwavering support in my musical (mis)adventures over the years, and to my close family and friends, who continually encourage me to keep exploring and growing.

\chapter*{Abstract}\label{chap:abstract}
\addcontentsline{toc}{chapter}{Abstract} 
Beat and downbeat tracking, jointly referred to as \textit{Meter Tracking}, is a fundamental task in Music Information Retrieval (MIR). Deep learning models have far surpassed traditional signal processing and classical machine learning approaches in this domain, particularly for Western (Eurogenetic) genres, where large annotated datasets are widely available. These systems, however, perform less reliably on underrepresented musical traditions.  

Carnatic music, a rich tradition from the Indian subcontinent, is renowned for its rhythmic intricacy and unique metrical structures \textit{(t\={a}\d{l}as)}. The most notable prior work on meter tracking in this context employed probabilistic Dynamic Bayesian Networks (DBNs). The performance of state-of-the-art (SOTA) deep learning models on Carnatic music, however, remains largely unexplored.  

In this study, we evaluate two models for meter tracking in Carnatic music: the \textit{Temporal Convolutional Network} (TCN), a lightweight architecture that has been successfully adapted for Latin rhythms, and \textit{Beat This!}, a transformer-based model designed for broad stylistic coverage without the need for post-processing. Replicating the experimental setup of the DBN baseline on the Carnatic Music Rhythm (CMR$_f$) dataset, we systematically assess the performance of these models in a directly comparable setting. We further investigate adaptation strategies, including fine-tuning the models on Carnatic data and the use of musically informed parameters.  

Results show that while off-the-shelf models do not always outperform the DBN, their performance improves substantially with transfer learning, matching or surpassing the baseline. These findings indicate that SOTA deep learning models can be effectively adapted to underrepresented traditions, paving the way for more inclusive and broadly applicable meter tracking systems.

\newpage

\tableofcontents
\newpage

\chapter{Introduction}

\section{Background}

Rhythm analysis is a central topic in Music Information Retrieval (MIR), aimed at computationally analysing or modelling the temporal structure of music. It encompasses a variety of tasks such as onset detection, tempo estimation, beat and downbeat tracking, pattern analysis, microtiming analysis and synchronization among others, which together enable a comprehensive understanding of musical timing.

This work focuses on the task of automatic estimation of beats and downbeats, commonly referred to as \textit{Meter Tracking}, critical for several higher-level MIR tasks such as music segmentation and structural analysis, as well as applications such as DJ mixing and automatic beat matching.

\subsection{Metrical Structure in Music}

\begin{figure}[htbp]
    \centering
    \includegraphics[width=0.6\textwidth]{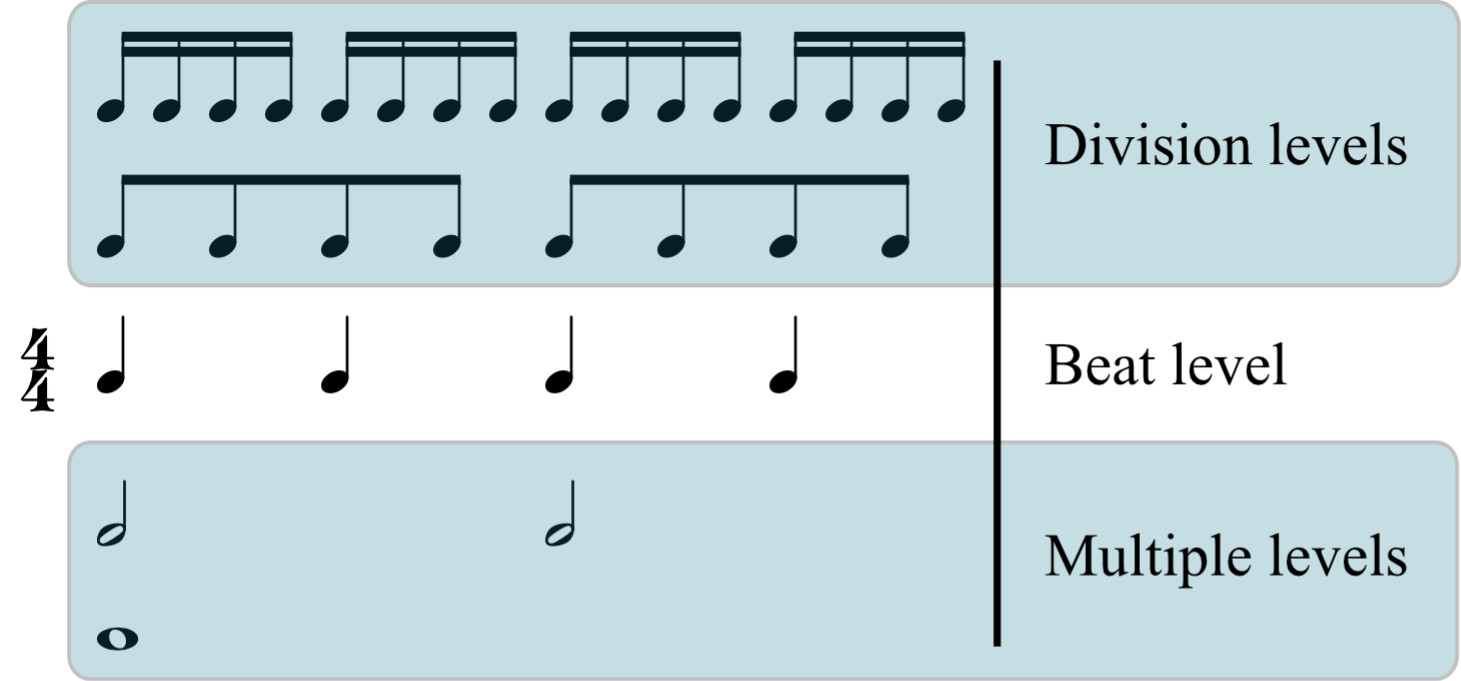}
    \caption[Musical meter in Western music]{Musical meter in Western music. Figure from Wikipedia, 
    \href{https://en.wikipedia.org/wiki/Metre_(music)}{Metre (music)}.}
    \label{fig:meter_music}
\end{figure}

Rhythm in music is perceived as pulsations organized at multiple hierarchical levels of differing timespans, known as its \textbf{meter} or \textbf{metrical structure} \cite{bilmes1993timing, London_hearingtime2012}. These levels range from very fast subdivisions to larger organizational units (for example, see figure \ref{fig:metrical_levels}). The different metrical levels are described as follows:

\begin{description}

\begin{figure}[htbp]
    \centering
    \includegraphics[width=\textwidth]{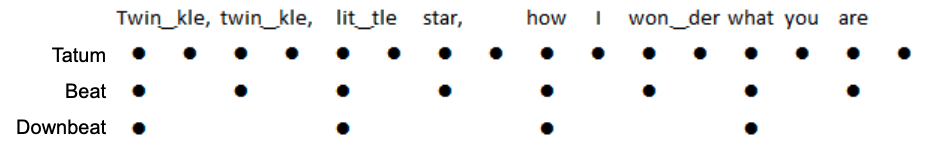}
    \caption[Example of perceived metrical levels]{Perceived metrical levels in '\textit{Twinkle, Twinkle, Little Star}'}
    \label{fig:metrical_levels}
\end{figure}

  \item[\textbf{Tatum}] 
  The fastest regular pulse in the music that listeners can perceive as a meaningful subdivision of rhythm. Often corresponds to a 16th note in Western music, but the specific duration depends on the tempo and style.

  \item[\textbf{Tactus} (Beat)] 
  The perceptually most salient pulse level that a listener would naturally tap their foot to. It typically corresponds to quarter notes in Western music but again depends on tempo and context. The beat level is central to most rhythm perception tasks and often serves as the reference level for tempo.

  \item[\textbf{Meter (Bar, Measure, Cycle)}] 
A  grouping of beats into a recurring structure that establishes musical phrasing and form. Measures are typically marked by accent patterns and serve to shape listeners’ expectations of timing and emphasis.

In Western music, meter is commonly represented using time signatures, such as 4/4, which indicates four beats per measure, with each beat typically being a quarter note in duration (see figure \ref{fig:meter_music}). Other common meters include 3/4 (e.g., waltz).

  \item[\textbf{Downbeat}] 
  The first beat of a bar or cycle, often marked by a strong accent or structural change. It acts as a temporal anchor and plays a crucial role in conveying the start of a measure. Accurate downbeat perception is essential for understanding musical form and phrasing.

\end{description}

\subsection{Rhythm in Carnatic Music}

Carnatic music, one of the two principal traditions of Indian art music (IAM), is predominantly practiced and appreciated in the southern regions of the Indian subcontinent. It is distinguished by its dedicated audiences, sophisticated theoretical framework, and high level of musicianship.  

Traditionally, training in Carnatic music is transmitted orally through a lineage of teachers, with a strong emphasis on performance and improvisation. A typical Carnatic performance features a lead performer, a rhythmic accompaniment, a continuous background drone, and a melodic accompaniment. Unlike Western tonal music, Carnatic music does not employ harmony; instead, it is structured around the melodic framework of \textit{rāga} and the rhythmic framework of \textit{tāḷa} \cite{sambamoorthy_south_1998}. Consequently, Carnatic music has become an important subject in MIR research as it presents unique challenges and opportunities for computational analysis \cite{rao2023indian}. 

Rhythmic organization in Carnatic music is governed by the tāḷa system, a hierarchical framework of time cycles that underlies melodic and rhythmic phrasing as well as improvisation. Within each tāḷa cycle, sub-structures are defined to track progression through the cycle. While there are some conceptual parallels with Western metrical structures, the terminology and organization within the tāḷa system differ significantly. Table \ref{tab:western_carnatic_comparison} presents approximate correspondences between metrical hierarchies in Western and Carnatic frameworks.

\begin{table}[h]
\centering
\begin{tabular}{ll}
\toprule
\textbf{Western} & \textbf{Carnatic} \\
\midrule
Tatum & \textit{akṣara} \\
Beat & (indicated by hand gestures) \\
Measure & \textit{āvartana} \\
Downbeat & \textit{sama} \\
\bottomrule
\end{tabular}
\caption{Mapping of Western and Carnatic rhythmic concepts}
\label{tab:western_carnatic_comparison}
\end{table}

Moreover, the tāḷa framework includes elements unique to the Carnatic tradition that lack direct equivalents in Western metrical theory, resulting in 175 theoretically possible tāḷas. In practice, however, a core set of 35 tāḷas is predominantly used in performance and pedagogy. Table \ref{tab:popular_talas} lists the four most commonly employed tāḷas in Carnatic music, alongside their total number of beats per cycle.

\begin{table}[h]
\centering
\begin{tabular}{lc}
\toprule
\textbf{Tāḷa} & \textbf{\#Beats} \\
\midrule
\textit{Ādi} & 8 \\
\textit{Rūpaka} & 3 \\
\textit{Miśra chāpu} & 7 \\
\textit{Khaṇḍa chāpu} & 5 \\
\bottomrule
\end{tabular}
\caption{Popular \textit{tāḷas} in Carnatic music}
\label{tab:popular_talas}
\end{table}

\begin{figure}[htbp]
    \centering
    \includegraphics[width=0.5\textwidth]{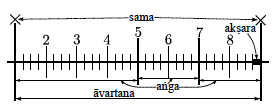}
    \caption[Illustration of Ādi tāḷa]{Illustration of Ādi tāḷa. Figure from \cite{srinivasamurthy_data-driven_2016}}
    \label{fig:aditala}
\end{figure}

Figure \ref{fig:aditala} illustrates some of the concepts from table \ref{tab:western_carnatic_comparison} using the example of an Ādi tāḷa cycle (8 beats). It also demonstrates how beats are further grouped into sections called \textit{aṅgas}. The progression through the tāḷa cycle is marked by distinctive hand gestures, which indicate both individual beats and the different types of sections within the cycle.

The parallels outlined above, though not exact, provide a conceptual bridge between Western metrical hierarchies and Carnatic tāḷa organization. This, in turn, enables the adaptation of meter-tracking algorithms originally developed for Western music to the context of Carnatic music.

\section{Motivation}

Over the past decade, deep learning approaches to meter tracking have significantly outperformed traditional signal processing and machine learning approaches. However, these models often excel primarily on Western (Eurogenetic) music genres, with limited generalisation to non-Western musical traditions due to several factors:

\begin{enumerate}
    \item \textbf{Data Representation and Bias}: Most existing deep learning-based methods utilise supervised learning and require large amounts of high-quality annotated training data. Available datasets are often dominated by Western music. Creating such datasets for non-Western music is a challenging and expensive process. It often requires culturally aware expertise to provide accurate annotations for beat and downbeat locations, particularly in music with complex meters.
    \item \textbf{Complexity of Rhythmic Structures}: Many non-Western music traditions possess complex and culture-specific rhythmic structures that differ significantly from those typically found in Western popular music. For instance, a large number of tracks in commonly used datasets primarily feature time signatures like 3/4 and 4/4. In contrast, non-Western music, such as IAM, utilise a variety of less common time signatures (e.g., 5/4 and 7/4) and metrical structures (tāḷa framework).    
    \item \textbf{Challenges in Downbeat Detection}: Even in the best-performing systems, downbeat detection tends to lag behind beat detection. Downbeat prediction often relies on harmonic or timbral cues, such as chord changes that typically coincide with downbeats. However, such correlations may be absent or less pronounced in many non-Western traditions.
\end{enumerate}

\subsection{Specific Challenges in Carnatic Music}

Carnatic music presents unique challenges for meter tracking systems. These challenges stem from the intricate nature of the rhythmic framework in the tradition and include:

\begin{itemize}
\item \textit{Diverse rhythmic features and practices}: The frequent use of improvisation, along with subtle variations in timing (microtiming), presents a challenge for systems relying on fixed rhythmic patterns. Syncopation, where beats are displaced from their expected positions, use of creative phase offsets (\textit{eḍupu}) and polyrhythms further complicates accurate meter tracking.
\item \textit{Percussion is not necessarily the timekeeper}: In a Carnatic performance, the rhythm is not solely determined by percussion instruments. The use of vocal cues (e.g., \textit{konnakol} - the art of vocal percussion) and hand gestures often takes on the role of guiding the rhythm. This can lead to an absence of clear, consistent percussion-based timekeeping.
\item \textit{Beats are not necessarily isochronous}: The definition of a beat as an isochronous pulse in the western paradigm does not extend directly to Carnatic music. Based on the tāḷa, the tactus or the foot tapping pulse can align either with the isochronous pulse or with the non-isochronous section (aṅga) markers. For the purposes of MIR methodologies, we adopt the former interpretation.
\end{itemize}

In light of these challenges, there is a clear opportunity to move beyond one-size-fits-all approaches towards adaptive approaches for meter tracking that account for the complexities of non-Western musical traditions such as Carnatic Music. Finally, the broader goal of this research is to bridge the gap between the state-of-the-art (SOTA) in computational rhythm analysis and the rich rhythmic expressions of musical cultures worldwide.

\section{Research Question and Objectives}
\subsection{Research Questions}

Based on the above discussion, the central research questions that guide this work are as follows:

\begin{quote}
\textit{To what extent do state-of-the-art deep learning models for meter tracking generalize to Carnatic music?}  
\end{quote}
\begin{quote}
\textit{Furthermore, is it possible to adapt these models to perform effectively on Carnatic music?}
\end{quote}


\subsection{Objectives}

In order to address the research questions, the following objectives are pursued:
\begin{enumerate}
\item \textbf{Benchmarking state-of-the-art models against a baseline}

Evaluate the performance of two state-of-the-art meter tracking models, \textit{Temporal Convolutional Network (TCN)} and \textit{Beat This!}, against a baseline set using Dynamic Bayesian Networks (DBN) by replicating the experimental setup employed in the baseline approach.

\textit{Rationale} : DBN has been widely used for meter tracking and has demonstrated effectiveness on Carnatic music. TCN and Beat This! are two distinct neural architectures that achieve leading performance on predominantly Western datasets. Comparing these approaches provides insights into the strengths and limitations of contemporary models when applied to non-Western music, thereby forming the groundwork for subsequent investigation.

\item \textbf{Exploring adaptive training strategies} 

Investigate different training strategies - namely, training models from scratch and fine-tuning pre-trained models - using Carnatic music data to assess their impact on performance.

\textit{Rationale} : This objective aims to evaluate whether exposure to domain-specific data enables existing models to better capture its unique rhythmic structures. Fine-tuning pre-trained models and training models specifically on Carnatic data offer promising adaptive strategies.

\item \textbf{Integrating musically informed features} 

Identify potential ways to integrate musically informed features, such as tāḷa structure and tempo information, into the system's parameters to enhance prediction accuracy.

\textit{Rationale} : By embedding musicological knowledge into computational frameworks, the system can account for culturally specific rhythmic patterns that strongly influence meter perception in Carnatic music. This integration seeks to bridge the gap between traditional theory and modern machine learning approaches.

\item \textbf{Performance analysis across tāḷa and tempo variations.}

Conduct a detailed evaluation of model performance across various tāḷas and tempo ranges, with a focus on identifying systematic strengths and weaknesses.

\textit{Rationale} : This analysis is essential for understanding where models succeed and where they fail in capturing rhythmic intricacies. The findings will highlight areas requiring targeted improvements, guiding future model development and helping to advance meter tracking for Carnatic music and other non-Western traditions.
\end{enumerate}

\chapter{State of the Art}\label{chap:sota}

This section traces the development of automatic beat and downbeat tracking and known applications to Carnatic music over the past decade. The structure follows the ISMIR 2021 tutorial on Tempo, Beat, and Downbeat Estimation \cite{tempobeatdownbeat:book}, but is updated to reflect more recent advances.

\section{Signal Processing Approach}\label{sec:signal_approach}

Traditional signal processing approaches to beat and downbeat tracking typically follow a two-stage process.

The first stage is \textbf{feature extraction}. Raw audio is first converted into a suitable time–frequency representation such as the Short-Time Fourier Transform (STFT), log or Mel spectrograms, and their variants. This is followed by conversion to a mid-level representation indicative of rhythmic events such as onsets - the instants marking the start of a note or a percussive stroke. The most common representation here is the Onset Detection Function (ODF), or \textit{novelty function}, which quantifies changes in signal parameters such as amplitude, energy, phase, and spectral content. A popular representation is the \textit{spectral flux}, which measures the positive change in spectral energy across frames.

\begin{figure}[htbp]
    \centering
    \includegraphics[width=\textwidth]{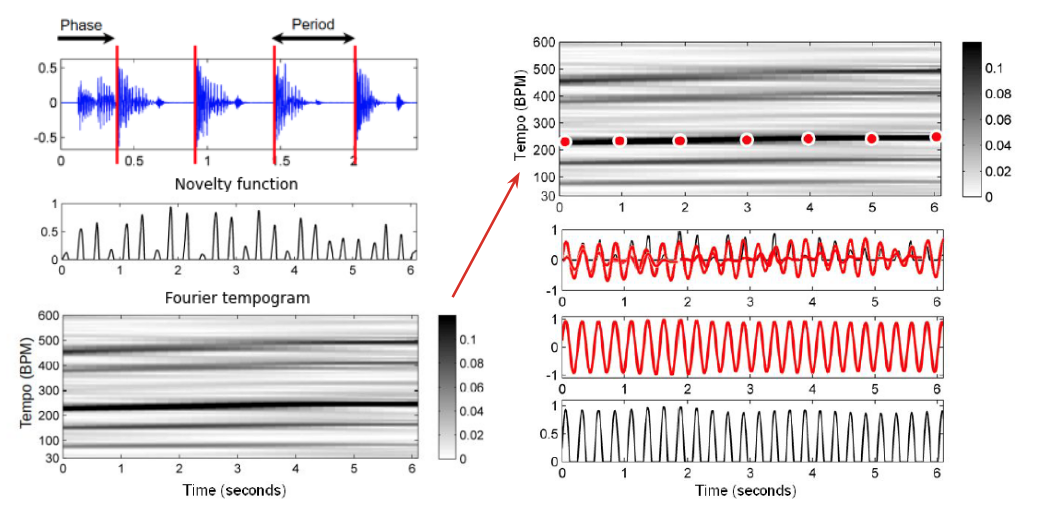}
    \caption[Predominant Local Pulse (PLP)]{Predominant Local Pulse (PLP). Figures taken from \cite{muller_fundamentals_2021}.}
    \label{fig:plp}
\end{figure}

The second stage is \textbf{periodicity estimation}. This typically involves autocorrelation or the computation of a Fourier Tempogram from the novelty function, where peaks indicate potential beat periodicities (i.e., tempo). Once a tempo estimate is obtained, the next step is beat sequence estimation. A widely used technique for this is the Predominant Local Pulse (PLP), which enhances local periodicities in the novelty function and then applies peak picking to extract a sequence of locally significant beats. Figure \ref{fig:plp} is a simplified illustration of the computation of PLP from the novelty function. For music with a stable tempo, dynamic programming techniques \cite{ellis_beat_2007} are also employed to determine a globally optimal sequence of beats under an assumed tempo. Chapter 6 from Meinard Müller’s book on music processing \cite{muller_fundamentals_2021} and the accompanying Python notebooks are excellent resources for detailed explanations of the techniques described above.

Downbeat tracking, the estimation of the beginning of a bar, was traditionally treated as a separate task, often performed after beat tracking. Classical approaches leverage musical knowledge, particularly the tendency in Western music for chord changes and harmonic cues to align with downbeats. Machine learning extensions, on the other hand, compute features at beat positions and train classifiers to identify downbeats, with dynamic programming commonly used to enforce a musically plausible downbeat sequence.

It is evident that the effectiveness of these traditional signal-processing approaches depends heavily on parameter tuning at each stage. They are also limited by factors such as weak onset strengths, tempo variations, expressive timing, and the complexity of rhythmic structures, such as in Carnatic music. Moreover, strategies effective for Western music often do not transfer smoothly to other musical traditions.

\section{Bayesian Approach}\label{sec:bayesian_approach}

The task of jointly tracking beats, downbeats and tempo is also known as meter tracking. This joint approach is useful because beats and downbeats are intrinsically linked. Beats occur at a periodic rate (tempo), and downbeats occur in relation to beats and bar structure. The natural relationship between them can be exploited for a more accurate and musically coherent analysis compared to treating them as separate tasks.

The Bayesian approach provides a probabilistic framework for jointly modelling hidden variables (such as beats and downbeats) from observations (novelty function) using:
$$
P(\text{hidden state} \mid \text{observations}) \propto P(\text{observations} \mid \text{state}) \cdot P(\text{state})
$$

A \textbf{Probabilistic Graphical Model} (PGM) provides a structured, visual representation of how random variables, both observed and hidden, are interrelated via their conditional dependencies \cite{koller_probabilistic_2009}. In a PGM, \textit{nodes} represent these variables, where observed variables are typically shown as shaded nodes and hidden variables as unshaded nodes and \textit{edges} denote probabilistic dependencies between them.
 
A \textbf{Dynamic Bayesian Network} (DBN) is a type of PGM that specifically models temporal processes \cite{Murphy2002}. Let the observed sequence be denoted:
$$
\mathbf{y}_{1:K} = \{y_1, y_2, \dots, y_K\}
$$
where $K$ is the number of frames. The goal is to infer hidden variables:
$$
\mathbf{x}_{1:K} = \{x_1, x_2, \dots, x_K\}
$$

The joint distribution over observations and hidden states factorizes as:
$$
P(y_{1:K}, x_{0:K}) = P(x_0) \prod_{k=1}^{K} P(x_k \mid x_{k-1}) P(y_k \mid x_k)
$$
where:\\
$P(x_0)$ is the prior over initial states,\\
$P(x_k \mid x_{k-1})$ is the transition model,\\
$P(y_k \mid x_k)$ is the observation model.

In the context of meter tracking, DBNs can jointly model the relationship between metrical positions (beats and downbeats) and observed rhythmic patterns extracted from input features such as onset strength or spectral flux. These models can also explicitly incorporate musical knowledge such as typical tempo ranges, metric structures and rhythmic patterns (as in the case of Tāḷas in Carnatic music) by setting the priors learned from the training data. 

\subsection{Bar Pointer model}\label{subsec:bar_pointer}

The \textbf{Bar Pointer Model} (BP-model) \cite{whiteley_bayesian_2006} is a specific type of DBN that has been successfully applied to meter tracking. Its effectiveness in inferring meter and rhythmic styles has been demonstrated on culturally diverse music corpora, including Ballroom dance music \cite{krebs_rhythmic_2013}, as well as Turkish, Greek, and Carnatic music \cite{holzapfel_tracking_2014}, showing robustness in handling complex and non-Western meters. This makes the bar pointer model an established and powerful computational approach for meter tracking.

The model also features prominently in Ajay Srinivasamurthy’s 2016 doctoral thesis \cite{srinivasamurthy_data-driven_2016} on automatic rhythm analysis of Indian Art Music, which remains the most comprehensive study of meter tracking in Carnatic music to date. The bar pointer model introduces a hypothetical "pointer" that moves through the metrical cycle and resets at the downbeat.

\begin{figure}[htbp]
    \centering
    \includegraphics[width=0.3\textwidth]{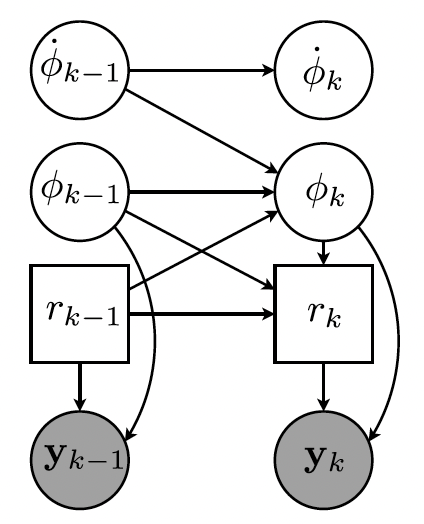}
    \caption[The Bar Pointer model]{The Bar Pointer model. Figure taken from \cite{srinivasamurthy_data-driven_2016}.}
    \label{fig:bar_pointer_model}
\end{figure}

\textbf{Hidden Variables in the BP-model}
\begin{itemize}
    \item Bar Position ($\phi_k$) : Variable indicating position in the bar; $\phi = 0$ denotes the downbeat.
    \item Tempo ($\dot{\phi}_k$) : Rate of progression of the pointer through the bar; modeled stochastically to allow for natural tempo fluctuation.
    \item Rhythmic Pattern Index ($r_k$) : Encodes discrete rhythmic templates, capturing expected accent structures across different metrical styles.
\end{itemize}

\textbf{Transition Model}

The transition model defines how the hidden state evolves over time:
$$
P(x_k \mid x_{k-1}) = P(\phi_k \mid \phi_{k-1}, \dot{\phi}_{k-1}, r_{k-1}) \cdot P(\dot{\phi}_k \mid \dot{\phi}_{k-1}) \cdot P(r_k \mid r_{k-1}, \phi_k, \phi_{k-1})
$$
The first term updates the bar position $\phi_k$ based on the previous position $\phi_{k-1}$ and tempo $\dot{\phi}_{k-1}$.
The second term enforces smooth tempo changes by modelling $\dot{\phi}_k$ based on $\dot{\phi}_{k-1}$.
The third term allows rhythmic pattern $r_k$ to change, but only at the end of a bar (i.e., when $\phi_k < \phi_{k-1}$).

\textbf{Observation Model}

The observation model $P(y_k \mid x_k)$ defines the likelihood of observing feature $y_k$ given the current state. It is often implemented using Gaussian Mixture Models (GMMs) trained on bar-position-aligned rhythmic patterns derived from annotated data. The model captures how likely an onset or spectral event is to occur at each position in the bar, for each pattern.

\subsection{Inference in Bayesian Meter Tracking}\label{subsec:inference_dbn}

Once a Bayesian model (like the bar pointer model) is defined, the core computational task is inference. Given the observations $y_{1:K}$, the goal is to estimate the hidden state sequence $x_{1:K}$ — tempo, bar position, and rhythmic pattern — that best explain the observed audio features. 
$$
\text{Goal: } \arg\max_{x_{1:K}} P(x_{1:K} \mid y_{1:K})
$$
Depending on how the hidden state space is modeled — discretely or continuously — different inference techniques are used. The two dominant approaches are:

\textbf{Viterbi Decoding}

The Viterbi algorithm is a dynamic programming method that finds the single most likely i.e. Maximum A Posteriori (MAP) sequence of hidden states. It assumes a discrete state space - the bar position $\phi$, tempo $\dot{\phi}$, and rhythmic pattern $r$ are discretized into a fixed grid.

This approach provides exact inference under the discrete model and is efficient when the state space is moderately sized. However, it becomes computationally expensive with fine discretization, for example in cases with long bars, and it is inflexible in real-time or online settings. To tackle these scalability challenges, Krebs et al. proposed an Efficient State Space Model \cite{krebs_efficient_2015} that restructures the original bar pointer model resulting in better accuracy and drastically reduced computational complexity.

\textbf{Particle Filtering}

When the hidden state space is modeled as continuous (or very high-dimensional), exact inference becomes intractable. Particle filtering provides an approximate solution by using a set of weighted samples, called particles, each representing a possible state trajectory. In other words, each particle represents a hypothesis for the hidden state at time $k$: $x_k^{(i)} = [\phi^{(i)}_k, \dot{\phi}_k^{(i)}, r^{(i)}_k]$.

Particle filtering naturally incorporates uncertainty and multimodality, such as multiple possible tempo hypotheses, making it better suited for online or real-time applications. However, it is computationally intensive, requires tuning the number of particles, and since it is an approximate method, the results may vary between runs.

\newpage
\section{Deep Learning Approach}\label{sec:dl_approach}

Currently, data-driven deep learning approaches dominate the landscape for the meter tracking task as they offer several advantages. Deep Neural Networks (DNN) are capable of learning complex representations from raw input data, processing large-scale datasets more efficiently, generalising better across data and multi-task learning. The availability of GPUs and specialized deep learning frameworks (e.g., TensorFlow, PyTorch) has made training and deploying DNNs more practical.

\subsection{DNN Pipeline for Meter Tracking}\label{subsec:dnn_pipeline}

\begin{figure}[htbp]
    \centering
    \includegraphics[width=\textwidth]{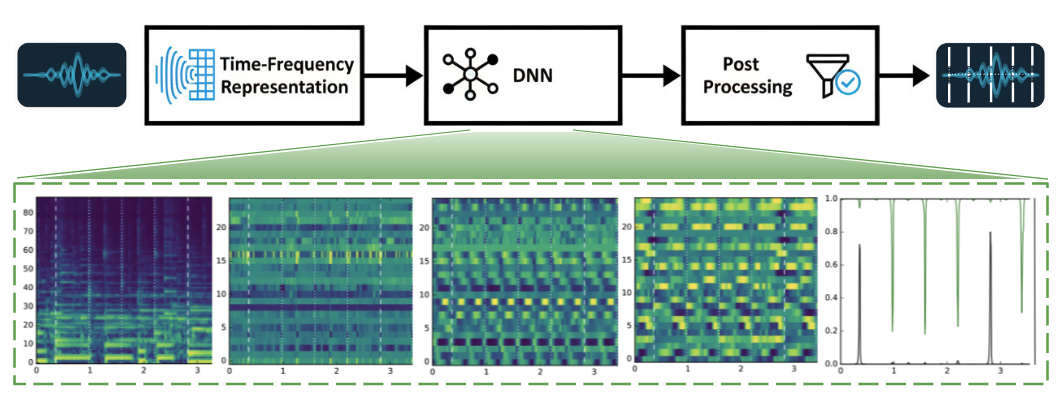}
    \caption[DNN based meter tracking pipeline]{DNN based meter tracking pipeline. Figure adapted from \textit{Tempo, Beat and Downbeat ISMIR Tutorial 2021} \cite{tempobeatdownbeat:book}.}
    \label{fig:dnn_pipeline_fig}
\end{figure}

A typical pipeline for a DNN-based meter tracking system (see Figure \ref{fig:dnn_pipeline_fig}) consists of two stages - \textbf{feature learning} and \textbf{temporal decoding}. DNNs first learn features from the input audio or its time-frequency representation and output an \textbf{activation} or salience function containing the possible beat and downbeat candidates. This is similar to the novelty function, but while the novelty function is derived from hand-crafted features, the activation function is produced by the network’s complex internal representation. 

The output activations of DNNs are often noisy and cannot be directly used for predictions. DBNs are proven in their ability to impose temporal consistency and metrical structure and are commonly used as a post-processing step to infer beats and downbeats from DNN activations. However, DBNs can also introduce several limitations due to their inherent properties. They do not work for music with time signature changes, tempo changes outside the prescribed range, and metric structures not represented in the state space. To overcome bias introduced by DBNs and generalise across music genres, recent efforts have attempted to remove this post-processing stage.

\subsection{Overview of Architectures}\label{subsec:architectures_overview}

Since beat and downbeat estimation is a sequence modelling problem, the most successful architectures applied to this task include Recurrent Neural Networks (RNNs), Temporal Convolutional Networks (TCNs) and transformer-based models, all of which are well-suited for capturing temporal dependencies in musical signals.

Böck et al. \cite{bock_joint_2016} utilise RNN, specifically Bidirectional Long Short-Term Memory (BLSTM) architecture, for a supervised classification task to simultaneously detect beats and downbeats. This significant work outperformed standalone DBN-based meter tracking, especially on the downbeat detection task for most Western music datasets.

Convolutional Neural Networks (CNN) are known to excel at extracting local features, such as transients, while having a relatively low model complexity. However, they suffer from a lack of long-term context, which makes it difficult to identify global rhythmic structures. Hybrid approaches that incorporate both spatial and temporal understanding are, therefore, utilised for meter tracking. BeatNet \cite{heydari_beatnet_2021} uses CRNN (Convolutional Recurrent Neural Network), which combines CNNs for feature extraction and recurrent layers for sequential modelling.

Temporal Convolutional Network has emerged as another powerful architecture for beat and downbeat tracking. TCNs utilise convolutional layers with dilations to achieve a large receptive field, allowing them to model long temporal contexts efficiently. 

More recently, the transformer architecture - originally successful in natural language processing - has been applied to meter tracking. Transformers utilise a self-attention mechanism that allows the model to weigh the importance of different parts of the input sequence when making predictions. This enables them to capture both local and global dependencies effectively while covering the entire input sequence. Hung et al. \cite{hung_modeling_2022} employ a spectral-temporal transformer (SpecTNT) architecture for this task. \textit{Beat This!}, a transformer-based system that removes the post-processing stage, achieves state-of-the-art beat and downbeat tracking performance on a number of Western music datasets. 

\section{Evaluation}\label{sec:evaluation}

The evaluation of meter tracking systems typically involves comparing predicted beat and downbeat times against annotated ground truth. In order to account for the inherent imprecision in annotations and musical events, most evaluation metrics allow a tolerance window around the annotated times.

\subsection{F-Measure}\label{subsec:fmeasure}

The F-measure, also known as the F1-score, evaluates the accuracy of predicted beat times by comparing them to ground truth annotations within a fixed temporal tolerance window (commonly ±70 ms). This metric intends to provide a measure of how many beats are correctly predicted without over or underpredicting. Downbeats are evaluated similarly, but due to their lower frequency, errors are more impactful. It is defined in terms of:

\textit{True Positives} ($N_{\text{TP}}$): Number of predicted beats that fall within the tolerance window of a ground-truth beat. \\
\textit{False Positives} ($N_{\text{FP}}$): Number of predicted beats that do not match any ground-truth beat within the tolerance window. \\
\textit{False Negatives} ($N_{\text{FN}}$): Number of ground-truth beats for which no predicted beat lies within the tolerance window.

\begin{figure}[htbp]
    \centering
    \includegraphics[width=0.7\textwidth]{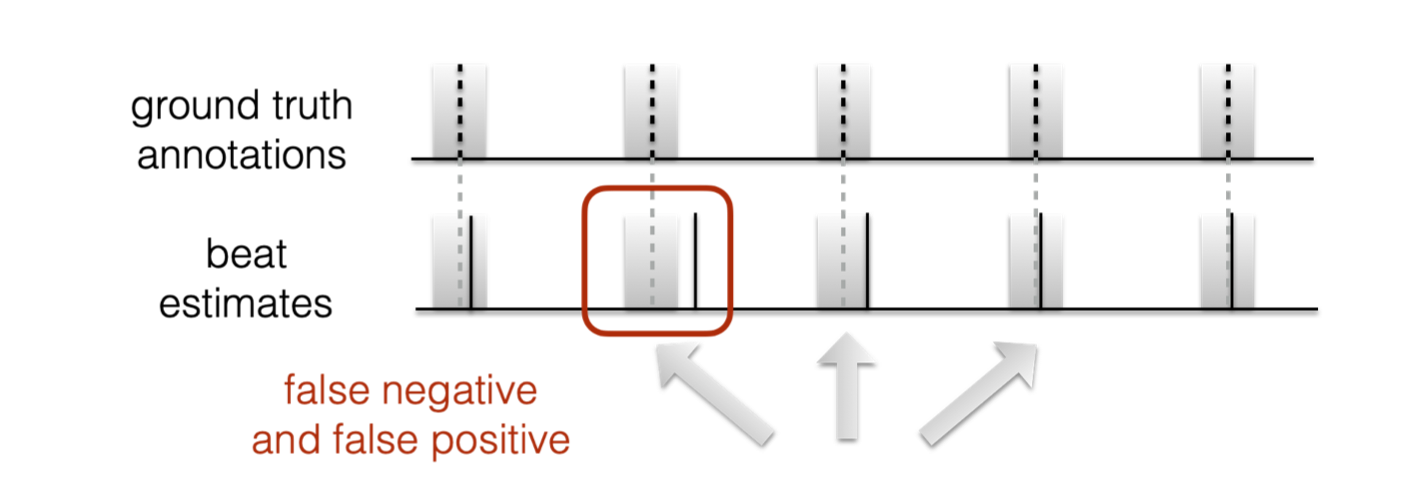}
    \caption[Tolerance window for F-measure]{Tolerance window for F-measure. Figure from \textit{Tempo, Beat and Downbeat ISMIR Tutorial 2021} \cite{tempobeatdownbeat:book}.}
    \label{fig:tolerance_window}
\end{figure}

Precision and Recall are defined as:
$$
\text{Precision} = \frac{N_{\text{TP}}}{N_{\text{TP}} + N_{\text{FP}}}, \quad
\text{Recall} = \frac{N_{\text{TP}}}{N_{\text{TP}} + N_{\text{FN}}}
$$

Then, the F-measure is the harmonic mean of precision and recall:
$$
F_1 = 2 \cdot \frac{\text{Precision} \cdot \text{Recall}}{\text{Precision} + \text{Recall}}
$$
While the F-measure is a widely used and intuitive metric, it is prone to systematic issues that can give a misleading impression of tracking quality. For instance, changing the size of the tolerance window can dramatically change the value of the measure. As a result of using a fixed tolerance window, beats inside the window are considered accurate regardless of their position inside the window. So, predictions consistently offset from the annotation would result a high F1 score. 

\subsection{Continuity-based Metrics}\label{subsec:continuity}

Continuity-based metrics were introduced to address some of these gaps by evaluating not just alignment accuracy, but also the consistency of metrical phase and tempo over extended regions. That is, evaluating not just whether beats are detected, but whether they are detected consistently across time and at the correct metrical level. This is especially important for applications such as real-time tracking, where maintaining stable and accurate beat information over time is crucial for synchronization and responsiveness.

\textbf{Continuity Criteria}

A predicted beat at time $\hat{b}_i$ is considered accurate only if it satisfies two conditions:

\begin{enumerate}
    \item The predicted beat $\hat{b}_i$ must lie within a predefined tolerance window around the corresponding ground-truth beat $b_i$. This window is not absolute but relative to the inter-beat interval (IBI), typically set to $\pm17.5\%$ of the local IBI.
    \item The preceding beat $\hat{b}_{i-1}$ must also fall within its own tolerance window. Furthermore, the IBI between $\hat{b}_{i-1}$ and $\hat{b}_i$ must be consistent with the IBI between $b_{i-1}$ and $b_i$.
\end{enumerate}

These conditions together define what is referred to as a \textbf{continuous segment}: a sequence of at least three consecutive beats that are temporally aligned, metrically consistent, and phase-correct. Only such segments contribute to the continuity-based metrics.

\textbf{Metrical Ambiguity}

Continuity metrics are designed to be sensitive to a range of metrical errors which may all have similar F-measure values but vastly different perceptual implications. To achieve this, continuity-based metrics introduce metrical variants of the reference annotation grid and evaluate predictions against each variant. The highest resulting score is selected. Commonly used metrical variants include:

\begin{itemize}
    \item Same metrical level, in-phase (i.e., beats align exactly with annotations)
    \item Same metrical level, off-phase (i.e., beats occur halfway between annotations)
    \item Double tempo (Twice the annotated metrical level)
    \item Half tempo (even-phase) (every other annotation starting from the first)
    \item Half tempo (odd-phase) (every other annotation starting from the second)
\end{itemize}

\textbf{Definitions of Continuity Metrics}

Let $N_{\text{correct}}^{\text{seg}}$ be the number of beats in the longest continuous correct segment, and $N_{\text{correct}}^{\text{all}}$ be the total number of correct beats (even across multiple segments). Four metrics are derived from this principle, distinguishing between strict (annotated) and lenient (allowed) metrical levels:

\newpage

\begin{itemize}
    \item CML$_c$(Correct Metrical Level - continuous):

$$
\text{CML}_c = \frac{N_{\text{correct}}^{\text{seg}}}{N_{\text{pred}}}
$$
    \item CML$_t$ (Correct Metrical Level - total):

$$
\text{CML}_t = \frac{N_{\text{correct}}^{\text{all}}}{N_{\text{pred}}}
$$
    \item AML$_c$ (Allowed Metrical Levels - continuous): 
    Same as CML$_c$, but allows metrical ambiguities.
    \item AML$_t$ (Allowed Metrical Levels - total):
    Same as CML$_t$, but allows metrical ambiguities.
\end{itemize}

Low continuity scores - especially when paired with a high F-measure - suggest that predictions are fragmented or metrically inconsistent, even if individual beats are frequently close to annotations. Comparing CML and AML variants can also reveal whether a system is making metrical-level errors (e.g., consistently tracking at half or double tempo) that still result in perceptually acceptable output. Overall, continuity metrics offer a more structurally aware evaluation than frame-level accuracy alone.

\chapter{DNN Models for Meter Tracking}\label{chap:architectures}

This work focuses on two main architectures: \textbf{Temporal Convolutional Network} and \textbf{Beat This!}. The following sections take a closer look at each system, explaining their key components and how they approach the task of meter tracking.

\section{Temporal Convolutional Network}\label{sec:tcn_arch}

\begin{figure}[htbp]
    \centering
    \includegraphics[width=0.5\textwidth]{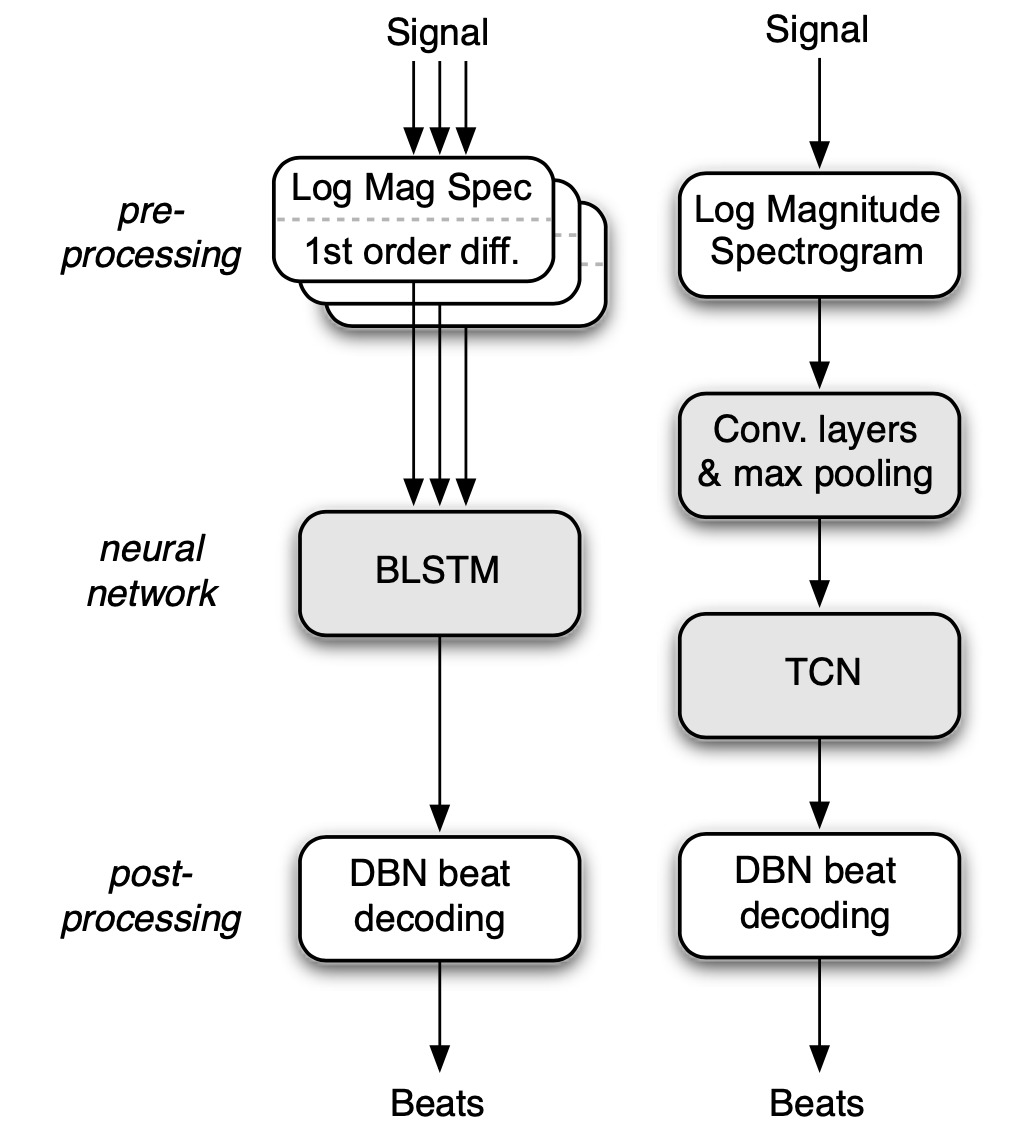}
    \caption[Comparison of BLSTM and TCN architectures]{Comparison of BLSTM and TCN architectures for beat tracking. Figure taken from \cite{matthewdavies_temporal_2019}.}
    \label{fig:lstm_vs_tcn}
\end{figure}

TCNs have been shown to outperform traditional RNN-based models such as BLSTMs in meter tracking tasks. The TCN architecture uses dilated convolutions to model temporal dependencies, allowing the model to process audio sequences in parallel. Unlike BLSTMs, which are inherently sequential and thus difficult to parallelize, TCNs enable parallel training across time steps, significantly reducing training times and computational costs. Through dilated convolutions, TCNs are capable of modelling long-range temporal dependencies (spanning entire bars or phrases) with significantly fewer parameters. These characteristics make TCNs not only more scalable but also better suited for real-time or low-latency applications. Figure \ref{fig:lstm_vs_tcn} shows an overview of beat tracking pipelines for the two architectures.

\subsection{Architectural Details}\label{subsec:tcn_archdetails}

There are two main components at the heart of a TCN-based meter tracker:

\textbf{Convolutional Block}

\begin{figure}[htbp]
    \centering
    \includegraphics[width=0.8\textwidth]{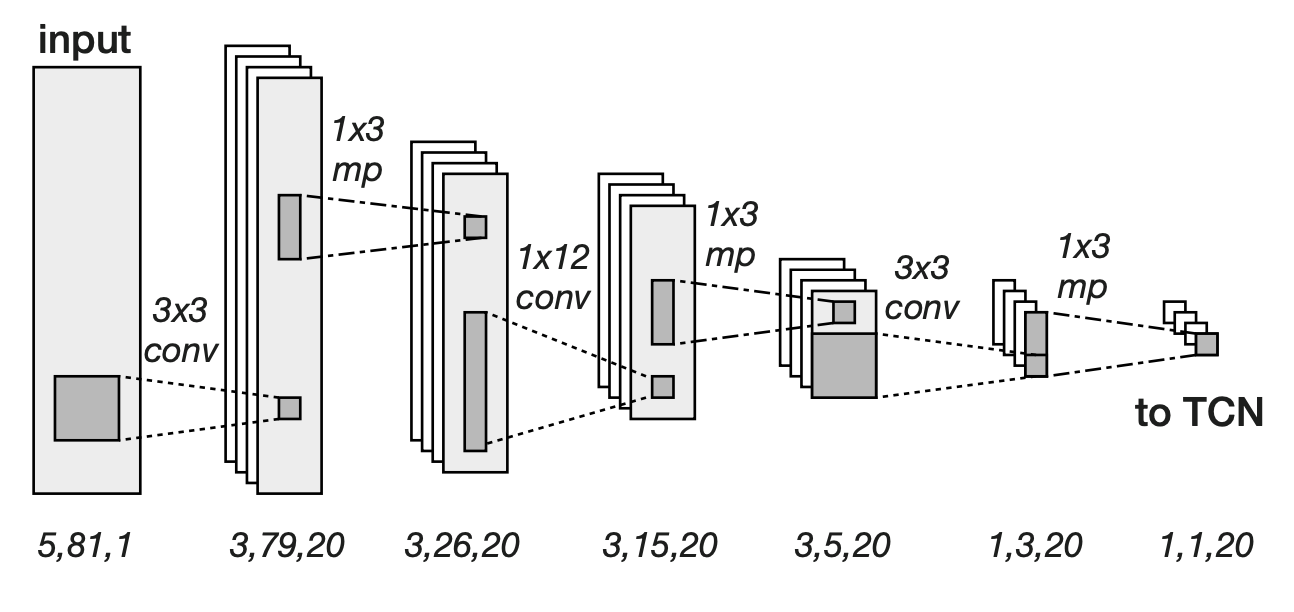}
    \caption[Convolutional block in a TCN-based meter tracker]{Convolutional block in a TCN-based meter tracker. Figure taken from \cite{bock_deconstruct_2020}.}
    \label{fig:tcn_conv}
\end{figure}

The convolutional block acts as the frontend feature extractor in the TCN-based meter tracking pipeline. Its role is to transform the input spectrogram into a more compact and informative set of learned features that emphasize the spectral-temporal patterns relevant to rhythm perception.
 Importantly, all convolution operations are performed without temporal downsampling. The convolutional block is designed to reduce spectral dimensionality while preserving the temporal resolution that is critical for tracking beat-related events. As seen in figure \ref{fig:tcn_conv}, a typical convolutional block includes:

\begin{itemize}
        \item Multiple 2D convolutional layers, each with a small kernel size (e.g., 3×3) to capture local time-frequency patterns.
        \item Pooling along the frequency axis, which compresses the spectral dimension while maintaining the original temporal resolution.
        \item Nonlinear activation functions, such as ELU, applied after each convolution to introduce nonlinearity.
    \end{itemize}

\textbf{TCN Block}

The input to the TCN is a highly sub-sampled feature vector derived from the magnitude spectrogram by the convolutional block, but which retains the same temporal resolution. The TCN block is the core temporal modelling component of the architecture. Its primary function is to model the sequential dependencies and periodic structures required for beat and downbeat prediction. It does so by learning filters via \textit{dilated} convolution. Dilation is equivalent to skipping samples in the input sequence. 

\begin{figure}[htbp]

    \centering
        \includegraphics[width=1\textwidth]{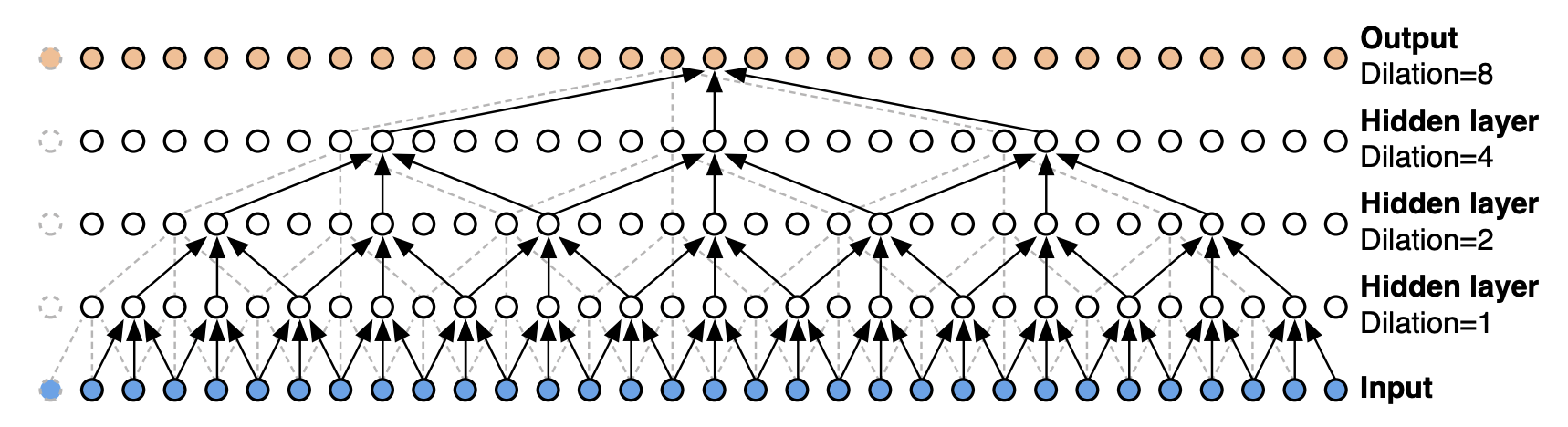}
    \caption[Temporal Convolutional Network]{Temporal Convolutional Network. Figure taken from ISMIR 2021 tutorial on Tempo, Beat, and Downbeat Estimation.\cite{tempobeatdownbeat:book}.}
    \label{fig:tcn_dilation}
\end{figure}

In a standard 1D convolution, each filter “slides” across the time axis of the input feature map, processing a local window (e.g., 3 frames) at each step. This is analogous to scanning for repeating rhythmic motifs. However, to model longer contexts, TCNs introduce dilated convolutions. A dilation defines the spacing between the elements in the filter's receptive field. For example, referring to figure \ref{fig:tcn_dilation} :

A dilation of 1 corresponds to adjacent time steps\( (t-1,\ t,\ t+1) \).\\
A dilation of 2 looks at every second time step \( (t-2,\ t,\ t+2) \).\\
A dilation of 4 expands further \( (t-4,\ t,\ t+4) \).

By stacking layers with exponentially increasing dilations (e.g., 1, 2, 4, 8...), the network can effectively model patterns over a large time span without a proportional increase in the number of parameters.

\textbf{Advantages for Meter Tracking:}
\begin{itemize}
    \item \textbf{Temporal resolution is preserved:} Unlike RNNs, TCNs can maintain the full temporal granularity of the input.
    \item \textbf{Efficient long-term modelling:} Due to dilation, a TCN with 10 layers and kernel size 3 can access $2^{10} = 1024$ time steps---several seconds of music---without loss of resolution.
    \item \textbf{Parallel training:} All time steps can be processed simultaneously, making the model highly suitable for GPU acceleration.
\end{itemize}

\subsection{Adaptation and Generalization}\label{subsec:tcn_adapt}

In the context of meter tracking, Davies and Böck \cite{matthewdavies_temporal_2019} first successfully repurposed the TCN design inspired by WaveNet \cite{van2016wavenet} for beat tracking. Building on this, they subsequently adopted a \textit{Deconstruct, Analyse, Reconstruct} approach \cite{bock_deconstruct_2020} to significantly improve performance by redesigning the TCN's convolutional and dilated layers, incorporating multi-task learning for tempo, beat, and downbeat estimation.

Research by \cite{maia_adapting_2022} further explored the adaptability of TCNs to non-Western and underrepresented music traditions, such as samba and candombe. Their approach used a TCN model trained initially on Western music and subsequently fine-tuned with limited annotations through transfer learning and data augmentation. The beat tracking F-measures exceeded 90\% with just 1.5 minutes of annotated training data, especially for more metrically homogeneous genres like candombe. This finding has important implications for ethnomusicologically-oriented MIR, where large annotated corpora are rarely available.

\subsection{Multi-task Learning Formulation}\label{subsec:tcn_multitask}

One of the key contributions of Davies and Böck's work was the use of a multitask learning paradigm within the TCN framework. Rather than training separate models for tempo, beat, and downbeat estimation, the system learns to predict multiple rhythmic targets simultaneously. This shared representation encourages generalization and enables the network to exploit the interdependence between tasks. In practice, this is implemented by:
\begin{enumerate}
    \item Using a \textbf{shared convolutional frontend and TCN backbone}, which learns a general representation of the rhythmic content.
    \item Adding separate \textbf{task-specific output heads} that predict:
    \begin{itemize}
        \item Beat Activation
        \item Downbeat Activation
        \item Tempo Estimation
    \end{itemize}
\end{enumerate}
Each output head is trained with its own loss function - typically binary cross-entropy loss for beats and downbeats - and the total loss is a weighted sum of these task-specific losses. This formulation encourages the model to learn both local and global rhythmic structure. Beats and downbeats are modeled jointly by framing the task as multi-label classification: for each input frame, separate binary classifiers determine the presence of a beat and/or a downbeat, allowing a frame to be labeled as either, both, or neither. 

Finally, the activations from the individual heads are then post-processed using DBN to yield discrete beat and downbeat position predictions.

\section{Beat This! : Tracker without Post Processing}\label{sec:beathis_arch}

In contrast to earlier meter tracking approaches that rely on DBNs for postprocessing, Beat This! \cite{foscarin_beat_2024} introduces a novel architecture that explicitly avoids such handcrafted temporal models, opting instead for a fully learnable and generalizable framework for beat and downbeat tracking. This architectural design seeks to overcome the need for explicit tempo or meter constraints, thus addressing known limitations of DBN-based postprocessing in handling expressive tempo changes, time signature shifts, and non-Western metrical structures. 

\subsection{Architectural Details}\label{subsec:beathis_archdetails}

\begin{figure}[htbp]
    \centering
        \includegraphics[width=0.5\textwidth]{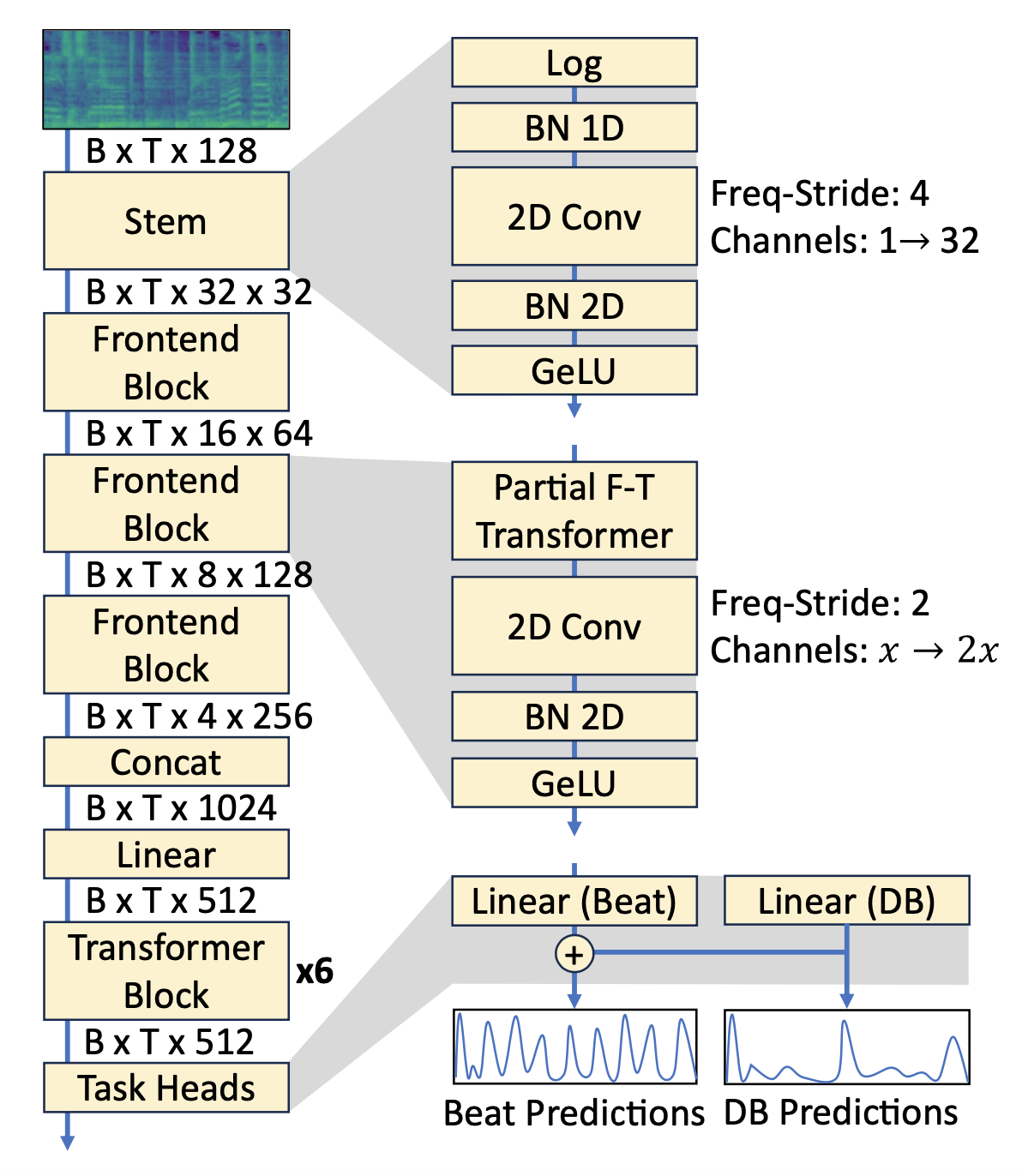}
    \caption[\textit{Beat This!} Architecture]{\textit{Beat This!} Architecture. Figure taken from \cite{foscarin_beat_2024}.}
    \label{fig:beat_this_arch}
\end{figure}

The Beat This! architecture shown in figure \ref{fig:beat_this_arch} can be broadly divided into three stages: a \textbf{frontend}, a \textbf{transformer-based sequence model}, and \textbf{task-specific heads} for prediction. 

\textbf{Frontend}

The audio is first converted into a mel spectrogram representation, which serves as the input to the model. The model processes a \textit{T}×128 spectrogram into
\textit{T}×2 probabilities; \textit{T} being the number of input frames. Processing begins with a \textit{Stem} block that standardizes the input across frequency bands and performs an initial convolution to extract low-level time-frequency patterns. This is followed by a series of blocks that combine convolutional layers with attention mechanisms, operating alternately over the frequency and time dimensions. Compared to the stacked convolution and pooling blocks typically used in TCN-based models, this design enables the model to integrate both timbral and rhythmic information early in the network.

\textbf{Transformer Backbone}

The sequence of features produced by the frontend is fed into a transformer encoder that that models temporal relationships using self-attention across time. Unlike TCNs, which model long-range dependencies through increasingly dilated convolutions, the transformer accesses global context at every layer. This is particularly advantageous for handling tempo changes, expressive timing, and other non-isochronous phenomena that are difficult to encode using fixed dilation patterns.

\textbf{Task Heads}

The final stage of the model consists of two task-specific output layers: one for beat prediction and one for downbeat prediction. Instead of treating the two outputs independently, the model introduces a \textit{Sum Head} mechanism encourages the downbeat predictions to reinforce beat predictions by summing the two outputs before computing the beat activation. This helps prevent inconsistencies, such as downbeats being predicted without corresponding beats, which can occur in architectures with separately trained outputs. The summed output is then used to produce the final beat probability sequence, while the downbeat output remains separate.

These activations are then passed to a lightweight postprocessing stage that extracts discrete beat and downbeat events without relying on tempo or meter constraints.

\subsection{Shift-tolerant Loss}\label{subsec:beathis_loss}

An important design element of the Beat This! model is its novel \textit{Shift-tolerant binary cross-entropy (BCE) loss}. In meter tracking, annotations are never perfectly precise. Even expert annotators disagree on exact beat placement due to performer expressiveness and perceptual ambiguity. A model trained with standard binary cross-entropy (BCE) loss penalises even slight temporal deviations from the annotated beat locations, which can lead to slow convergence and blurred activation peaks during training.
\[
\mathcal{L}_{\text{bce}}(y, \hat{y}) = -\sum_t \left[ y_t \log(\hat{y}_t) + (1 - y_t) \log(1 - \hat{y}_t) \right]
\]
TCN-based models, such as the one in \cite{bock_deconstruct_2020}, tackle this issue with \textit{target widening} - modifying the ground-truth labels by assigning partial positive weights to the frames surrounding each annotated beat (e.g., 0.5 for ±1 frame, 0.25 for ±2 frames). This helps to address the slow convergence but doesn't mitigate the blurred predictions.

The shift-tolerant loss addresses the latter issue by introducing a margin of temporal flexibility during loss computation, allowing the model to focus on producing sharp, confident peaks near annotated events, rather than learning widened targets.
\[
\mathcal{L}_{\text{st}}(y, \hat{y}, w) = -\sum_t \left[ w y_t \log\left(m_7(\hat{y})_t\right) + \left(1 - m_{13}(y)_t\right) \log\left(1 - m_7(\hat{y})_t\right) \right]
\]

Here, \( m_k(\cdot) \) denotes max-pooling over \( k \) frames. Positive labels are matched to the strongest nearby prediction (within 7 frames), and the loss ignores negative samples close to any label (13-frame window) to avoid penalizing valid near misses. Since beat/downbeat events are sparse in time compared to non-beat frames, a weight \textit{w}>1 is applied to the positive frames in the loss to ensure the model gives sufficient attention to correctly predicting these rare events. Beat This! sets \textit{w} to the ratio of negative to positive frames in the training set ensuring that both classes contribute approximately equally to the total loss.

\section{Practical Considerations: TCN vs Beat This!}\label{sec:final_comparison}

This section summarizes the key differences between the TCN and Beat This! in terms of data requirements, model complexity, resource usage, and meter tracking performance.

\textbf{Training Data Requirements:}
TCN-based systems have demonstrated strong performance with relatively modest training data, especially when augmented. In contrast, Beat This! relies heavily on extensive and diverse training data (over 4000 tracks from 18 datasets) and multiple augmentation strategies to further extend the training data. This larger training corpus is necessary to compensate for the absence of DBN-based postprocessing and to ensure generalizability across genres.

\textbf{Model Complexity and Parameters:}
The TCN model used by Davies et al. \cite{bock_deconstruct_2020} for multitask learning contains around 116k parameters. Beat This!, by comparison, is a significantly larger model (~20M parameters).

\textbf{GPU Usage and Efficiency:}
TCN architectures are lightweight and highly parallelizable, making them efficient to train even on modest GPUs or CPUs. Beat This!, while GPU-efficient, still requires more compute resources and memory, especially for data preprocessing and training (4 hours on A100).

\textbf{Performance Comparison:}
Comparing the performance of both the approaches on the GTZAN dataset (table \ref{tab:gtzan_comparison}), we see that Beat This! achieves state-of-the-art F1 scores without DBN postprocessing. On the downside, Beat This! sacrifices continuity for sharp, confident predictions. TCN with post-processing, while achieving comparable F1 scores, outperforms on continuity metrics. 

\begin{table}[h]
\centering
{\small
\resizebox{0.8\textwidth}{!}{%
\begin{tabular}{lcc|cc}
\toprule
\textbf{Model} 
  & \multicolumn{2}{c|}{\textbf{Beat}} 
  & \multicolumn{2}{c}{\textbf{Downbeat}} \\
\cmidrule(lr){2-3} \cmidrule(lr){4-5}
  & \textbf{F-measure} & \textbf{AMLt} 
  & \textbf{F-measure} & \textbf{AMLt} \\
\midrule
\textit{Beat This!} & \textbf{89.1 ± 0.3} & 89.8 ± 0.4 & \textbf{78.3 ± 0.4} & 79.1 ± 0.6 \\
TCN + DBN            & 88.5      & \textbf{93.1}  & 67.2      & \textbf{83.2}      \\
\bottomrule
\end{tabular}
} 
} 
\caption{Comparison of TCN and Beat This! performance on the GTZAN dataset.}
\label{tab:gtzan_comparison}
\end{table}

\chapter{Methodology}\label{chap:methodology}

This chapter outlines the methodology adopted for this study, detailing each step in the evaluation of meter tracking models for Carnatic music. It begins by describing the Carnatic Music Rhythm dataset ($\mathrm{CMR}_f$) used throughout the study, followed by an explanation of the baseline method, which this work builds on. Particular attention is paid to the experimental setup to ensure a fair and consistent comparison with the baseline and between models. Additionally, the chapter highlights musically informed techniques integrated into the training and post-processing stages (where applicable), aiming to adapt the models to the distinctive characteristics of the Carnatic music repertoire.

\section{Dataset}\label{sec:dataset}

The Carnatic Music Rhythm dataset ($\mathrm{CMR}_f$) \cite{srinivasamurthy2014supervised} is a rhythm-annotated corpus of Carnatic music performances. The dataset comprises audio excerpts with manually annotated time-aligned markers that track the progression through the tāḷa cycle. Along with the audio, the dataset includes tāḷa-related metadata, specifying the particular tāḷa used in each piece and other relevant information. A detailed summary of the dataset, including the four tāḷas and the number of pieces for each, is provided in Table \ref{tab:cmr_summary}.

\begin{table}[h]
\centering
{\small
\resizebox{0.9\textwidth}{!}{%
\begin{tabular}{lccc|c|c}
\toprule
\textbf{Tāḷa} & \textbf{\#Pieces} & \shortstack{\textbf{Total Duration} \\ \textbf{hours (min)}} & \textbf{Median Time} & \textbf{\#Ann.} & \textbf{\#Sama} \\
\midrule
Ādi (8)             & 50  & 4.21 (252.78) & 4m51s  & 22793  & 2882  \\
Rūpaka (3)          & 50  & 4.45 (267.45) & 4m37s  & 22668  & 7582  \\
Mīśra chāpu (7)     & 48  & 5.70 (342.13) & 6m35s  & 54309  & 7795  \\
Khaṇḍa chāpu (5)    & 28  & 2.24 (134.62) & 4m25s  & 21382  & 4387  \\
\midrule
\textbf{Total}   & 176 & 16.61 (996.98) & 5m04s  & 121602 & 22646 \\
\bottomrule
\end{tabular}
} 
} 
\caption{Summary of the Carnatic Music Rhythm dataset ($\mathrm{CMR}_f$)}
\label{tab:cmr_summary}
\end{table}

The $\mathrm{CMR}_f$ dataset contains pieces in four widely performed tāḷas that cover a majority of contemporary Carnatic music performances. These pieces include a mix of vocal and instrumental recordings from a wide variety of musical forms. Each piece is accompanied by percussion, mainly the mridangam. The collection includes pieces with a wide range of cycle durations, from less than one second to over 7 seconds. Each annotation in the dataset is a time-stamped marker that corresponds to a specific metrical position within the tāḷa cycle, including the sama (downbeat) and other beats. 

The ($\mathrm{CMR}_f$) dataset is available through the \href{https://compmusic.upf.edu/carnatic-rhythm-dataset}{CompMusic Project}. It is also supported by the \href{https://github.com/mir-dataset-loaders/mirdata}{\textit{mirdata}} Python library \cite{bittner_fuentes_2019} of dataloaders for various MIR datasets.

\section{Baseline}\label{sec:baseline_bpmodel}

The baseline for this study is based on the results obtained by \cite{srinivasamurthy_data-driven_2016} in his work on meter tracking using the $\mathrm{CMR}_f$ dataset and the Bar Pointer Model (refer to \ref{subsec:bar_pointer}). Table \ref{tab:bar_pointer_results} presents a summary of the meter inference results i.e., the uninformed meter tracking task using two different inference strategies, as obtained from the publication.

\begin{table}[h]
\centering
{\small
\resizebox{0.9\textwidth}{!}{%
\begin{tabular}{lcccc}
\toprule
\textbf{Acronym} & \textbf{Inference Algorithm} & \shortstack{\textbf{Beat} \\ \textbf{F-measure}} & \shortstack{\textbf{Beat} \\ \textbf{AML$_t$}} & \shortstack{\textbf{Downbeat} \\ \textbf{F-measure}} \\
\midrule
BP-HMM & Viterbi Algorithm & 0.718 & 0.722 & 0.44 \\
BP-AMPF & Auxiliary Mixed Particle Filter & 0.825 & 0.906 & 0.574 \\
\bottomrule
\end{tabular}
} 
} 
\caption{Baseline meter tracking performance on $\mathrm{CMR}_f$ using Bar Pointer model}
\label{tab:bar_pointer_results}
\end{table}

\subsection{Baseline Setup}\label{subsec:baseline_setup}

Described below are the specific components and methodology used in the original which will be replicated in our work.
\begin{enumerate}
    \item \textbf{Cross-Validation:}
    The setup involves a \textbf{two-fold cross-validation} procedure. The dataset is divided into two distinct folds for the cross-validation process.
    \item \textbf{Pre-determined Folds:}
    The setup uses \textbf{pre-determined train and test folds} ensuring that there is no overlap between the data within each fold. Each fold is constructed ensuring that both the training and testing datasets contain an \textbf{equal number of pieces}.
    \item \textbf{Number of Runs:}
    To account for the inherent variability in probabilistic modeling, the original study performs \textbf{three runs} of the cross-validation process. The final performance metrics are averaged over the three runs to obtain a mean performance score
    \item \textbf{Tāḷa Contribution:}
    The training data is \textbf{pooled from all tāḷas} in the dataset, ensuring that the model is exposed to a diverse range of rhythmic patterns. 
    \item \textbf{Preservation of Tāḷa Distribution:}
    The experimental setup \textbf{preserves tāḷa distribution} in each fold. Specifically, the distribution of tāḷas in the training and test folds mirrors the distribution of tāḷas in the full dataset. 
\end{enumerate}

\section{Experiment Setup}\label{sec:experimental_setup}

For both models under evaluation, we first replicate the data splits and training setup as described in the previous section. Additionally, we establish common procedural guidelines to ensure a fair and consistent comparison between the models:

\begin{itemize}
\item The dataset, comprising 176 samples, is divided into two predetermined folds of 88 examples each, identical to those used in the baseline experiment’s two-fold cross-validation scheme. In each iteration, one fold is used for training while the other serves as the test set, with the folds alternating roles between iterations.
\item The train fold is further subdivided into training (80\%) and validation (20\%) subsets. Consequently, each fold contains 70 training examples and 18 validation examples.
\item We perform three training runs per fold. To ensure reproducibility of validation splits and network initializations, we set predetermined random seeds [42, 52, 62] for each respective run. As a result, six distinct models are generated for every training strategy, and the results are reported as the mean performance across these six models.
\item Validation loss is employed as the primary metric for monitoring training progress and for early stopping. Training is terminated when no improvement in validation loss is observed.
\item The models are evaluated using the F-measure as well as the continuity metrics {CML$_t$} and {AML$_t$} for both beat and downbeat. The evaluation process is carried out using the Python package \href{https://github.com/mir-evaluation/mir_eval}{\textit{mir\_eval}} \cite{Raffel2014mir_eval}.
\end{itemize}

\subsection{TCN}\label{subsec:tcn_setup}

\textbf{Model}

The experimental setup for the TCN employed in this study is based on the open-source implementation of \textit{Deconstruct, Analyse, Reconstruct} \cite{bock_deconstruct_2020} made available by the authors as part of the \href{https://tempobeatdownbeat.github.io/tutorial/intro.html}{ISMIR 2021 tutorial on Tempo, Beat and Downbeat Estimation}. This implementation was subsequently repurposed in \textit{Adapting Meter Tracking Models to Latin American Music} \cite{maia_adapting_2022}, and an updated, user-friendly version is provided in the \href{https://lamir-workshop.github.io/lamir_hackathon/intro.html}{Tutorial for LAMIR 2024 Hackathon} \cite{morais_lamir_2024}. The present study utilises these prior works and their respective experimental setups as the basis for the TCN implementation.

\textbf{Trainable Parameters} : 72.3K

\textbf{Training Strategies}

For the TCN model, we evaluate three strategies inspired by \cite{maia_adapting_2022}.
\begin{itemize}
    \item \textit{Baseline (TCN-BL)}\\
    First, we train a model on the popular Western \href{https://ismir.net/resources/datasets/}{datasets for meter tracking}- GTZAN, Ballroom, Beatles and RWC datasets following \cite{maia_adapting_2022}. This model is assumed to be a good starting point for a baseline evaluation of the TCN on Carnatic data as well as for subsequent transfer learning experiments. Following protocol, we perform three training runs and report mean performance on the $\mathrm{CMR}_f$ dataset.
    \item \textit{Fine-tuning (TCN-FT)}\\
    Under this strategy, the model with the lowest validation loss from TCN-BL is used as a starting point for fine-tuning the network on Carnatic data. The assumption is that, although the model was pre-trained on Western datasets, it has learned a representation that can be adapted for a different musical tradition, as demonstrated in \cite{maia_adapting_2022}.
    \item \textit{Training from Scratch (TCN-FS)}\\
    This strategy involves training a randomly initialized network (using one of the predefined seed values) from scratch on each fold.
\end{itemize}

\textbf{Loss Function}\\
We employ a simple loss function defined as the sum of the binary cross-entropy (BCE) for beat and downbeat predictions:
$$
\mathcal{L} = \text{BCE}_{\text{beat}} + \text{BCE}_{\text{downbeat}}
$$

Table \ref{tab:tcn_settings} provides a summary of the training configuration settings employed across the different TCN strategies.
\begin{table}[h]
\centering
{\small
\resizebox{\textwidth}{!}{%
\begin{tabular}{lccccccc}
\toprule
\textbf{Acronym} & \textbf{Strategy} & \shortstack{\textbf{Models} \\ \textbf{Trained}} & \textbf{Epochs} & \shortstack{\textbf{Early} \\ \textbf{Stoppage}} & \shortstack{\textbf{Learning} \\ \textbf{Rate}} & \shortstack{\textbf{LR} \\ \textbf{Reduction} \\ \textbf{(Factor)}} \\ \midrule
TCN-BL & TCN Baseline & 3 & 100 & 20 & 0.005 & 0.2 \\
TCN-FS & Train from Scratch & 6 & 100 & 20 & 0.005 & 0.2 \\
TCN-FT & Finetune from Baseline & 6 & 50 & 10 & 0.001 & 0.2 \\
\bottomrule
\end{tabular}
} 
} 
\caption{Training configurations for the TCN strategies}
\label{tab:tcn_settings}
\end{table}

\textbf{Post Processing}

For post-processing network activations, a DBN-based post-processor is used. The Python library \href{https://github.com/CPJKU/madmom}{\textit{madmom}} \cite{madmom} offers an open-source joint beat and downbeat DBN post-processor approximated by a Hidden Markov Model (HMM), based on \cite{bock_joint_2016, krebs_efficient_2015}. In this work, we use its offline mode utilising the Viterbi algorithm for inference.

\newpage
\subsection{Beat This!}\label{subsec:beathis_setup}

\textbf{Model}

For Beat This!, we use the stock implementation for baseline evaluation. For fine-tuning, however, we build upon a \href{https://github.com/smilo7/more-beats-for-this}{modified implementation} that facilitates fine-tuning of the stock model. Despite these modifications, we retain the default training configurations, including data augmentation schemes, and utilise the pre-trained models provided by the original authors.

\textbf{Trainable Parameters} : 20.3M

\textbf{Training Strategies}

We adopt only two strategies: Baseline and Fine-tuning. Given that Beat This! is a transformer-based architecture, it is highly data-intensive, which makes training from scratch on a single dataset impractical. 

\begin{itemize}
    \item \textit{Baseline (BeatThis-BL)}\\
    We utilise the three pre-trained checkpoints provided with the stock model, namely \textit{final0}, \textit{final1} and \textit{final2}, all of which have been trained on a large corpus comprising 18 different datasets (excluding GTZAN). We evaluate these models on the $\mathrm{CMR}_f$ dataset and report the mean performance across all three models as the baseline performance for Beat This!.
    \item \textit{Fine-tuning (BeatThis-FT)}\\
    The fine-tuning process begins with the default (\textit{final0}) checkpoint as the pre-trained baseline, which is then further fine-tuned over the course of 50 epochs.
\end{itemize}

Lastly, we use the built-in \textit{Shift-tolerant weighted BCE} loss and skip post-processing.

\section{Musically Informed Strategies}\label{sec:culture_informed}

Incorporating musicological insights into meter tracking systems can enhance their performance. This section explores strategies used during training and post-processing to help our DNN meter tracking systems adapt better to Carnatic Music.

\subsection{Music-Informed Training}\label{subsec:informed_training}
These strategies applied at the training stage aim to ensure that the model is exposed to rhythmic diversity consistently during the training process:

\textbf{Stratified Tāḷa-based Train/Validation Split}\\
To ensure consistent performance across tāḷas, we implement a stratified train/validation split based on tāḷas. This balances representation of all tāḷas in training and validation, enabling per-tāḷa error analysis and targeted improvements for underperforming tāḷas.

\textbf{Interleaved Train Data Loader}\\
An issue with the $\mathrm{CMR}_f$ dataset is imbalanced class distribution - some tāḷas appear more frequently than others. We use an interleaved data loader that proportionally spaces rarer tāḷas like khaṇḍa chāpu in training, ensuring a more balanced learning process.

\subsection{Music-Informed Post-Processing}\label{subsec:informed_post}
\begin{figure}[htbp]
    \centering
    \includegraphics[width=0.85\textwidth]{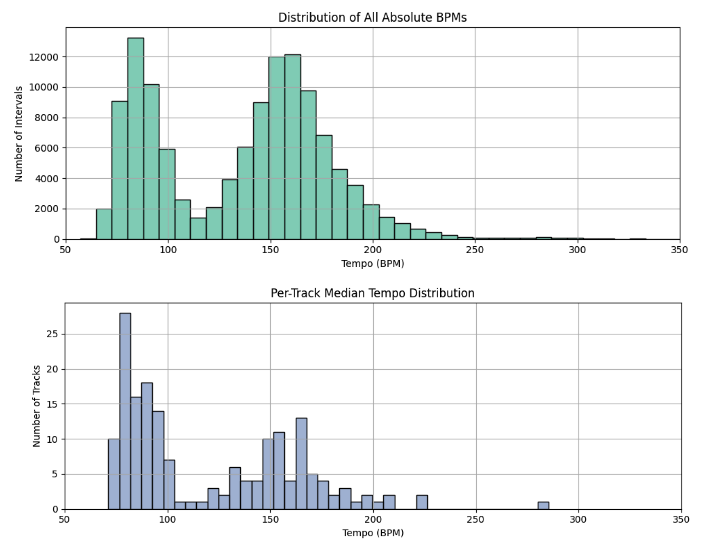}
    \caption[Distribution of tempos in the $\mathrm{CMR}_f$ dataset]{Distribution of tempos in the $\mathrm{CMR}_f$ dataset.}
    \label{fig:tempo_distribution}
\end{figure}

One impactful strategy for enhancing the performance of meter tracking systems is post-processing. The \href{https://madmom.readthedocs.io/en/v0.16/modules/features/downbeats.html}{DBNDownBeatTrackingProcessor} from the \textit{madmom} library allows us to tune parameters to reflect musical characteristics of the data being processed. We set the following parameters based on musicological knowledge as well as insights from the dataset:

\begin{itemize}
    \item \texttt{beats\_per\_bar} = [3, 5, 7, 8] based on the four tāḷas in our dataset instead of the default [3, 4]
    \item \texttt{min\_tempo} = 55 and \texttt{max\_tempo} = 230, reflecting the tempo range observed in the dataset (see Figure \ref{fig:tempo_distribution}). This range, chosen based on preliminary experiments, covers 99\% of all tempos and provides a constrained search space for the post-processor, with results showing slight performance improvement compared to a max tempo of 300 BPM (99.9\% of all tempos).
\end{itemize}

For reproducibility, all relevant code repositories, software resources, and dataset references utilised in the experiments are catalogued in Appendix~\ref{app:supplementary}.

\chapter{Results and Discussion}\label{chap:results}
This chapter presents and examines the performance of models trained with the various strategies described in the previous chapter. Each approach is evaluated using quantitative metrics, including F-measure and continuity scores, to provide a robust comparison. In addition, a detailed breakdown by tāḷa is conducted to highlight how each model responds to the unique rhythmic structures of Carnatic music, allowing their respective strengths and weaknesses to emerge more clearly.

\section{Model-wise Performance}\label{sec:model_performance}

Table \ref{tab:model_wise_performance} below shows the overall performance of the two models and their strategies against the Bar Pointer model baseline with the highest performing metrics highlighted in bold.

\begin{table}[h]
\centering
{\small
\resizebox{0.9\textwidth}{!}{%
\begin{tabular}{lccc|ccc}
\toprule
\textbf{Model} 
  & \multicolumn{3}{c|}{\textbf{Beat}} 
  & \multicolumn{3}{c}{\textbf{Downbeat}} \\
\cmidrule(lr){2-4} \cmidrule(lr){5-7}
  & \textbf{F-measure} & \textbf{{CML$_t$}} & \textbf{{AML$_t$}} 
  & \textbf{F-measure} & \textbf{{CML$_t$}} & \textbf{{AML$_t$}} \\
\midrule
BP-HMM          & 71.8 & ---  & 72.2 & 44.0 & ---  & ---  \\
BP-AMPF         & 82.5 & ---  & 90.6 & 57.4 & ---  & ---  \\
\midrule
TCN-BL          & 77.1 & 51.6 & 77.9 & 28.9 & 21.6 & 33.8 \\
TCN-FT          & 80.7 & 50.2 & \textbf{91.9} & 52.9 & 35.3 & 57.8 \\
TCN-FS          & 84.6 & 62.9 & 88.0 & 63.9 & \textbf{52.1} & \textbf{67.0} \\
\midrule
BeatThis-BL     & 71.3 & 39.2 & 56.8 & 27.6 & 2.0  & 8.7  \\
BeatThis-FT     & \textbf{90.3} & \textbf{78.0} & 80.0 & \textbf{66.8} & 38.2 & 53.7 \\
\bottomrule
\end{tabular}
} 
} 
\caption{Model-wise Performance Comparison}
\label{tab:model_wise_performance}
\end{table}

Both the TCN-BL and BeatThis-BL models, which were trained on Western music datasets, fail to achieve baseline performance levels in meter tracking for Carnatic music. Although the beat tracking accuracy of these models approximates baseline performance, their downbeat tracking performance remains substantially below baseline, despite being trained on extensive datasets. 
    
This disparity reveals the fundamental differences in rhythmic structures between Western and Carnatic music and illustrates the challenges faced by neural networks in directly transferring learned knowledge across distinct musical traditions.

In contrast, the TCN-FT model nearly attains baseline performance, notably achieving the highest beat {AML$_t$} score among all evaluated models. Interestingly, preliminary experiments demonstrated comparable results when fine-tuning a model initially trained only on the GTZAN dataset. 
    
These findings further highlight the necessity for the network to re-optimize its hyperparameters when adapting from Western to Carnatic music, indicating that the quantity of data used during pretraining may be less significant compared to the subsequent fine-tuning on Carnatic music. In fact, the performance of the TCN-FT model may be hindered by being undertrained. Additional fine-tuning could enhance the results, effectively equating to training the model from scratch.

Both TCN-FS and BeatThis-FT significantly outperform the DBN baseline in beat and downbeat tracking, with BeatThis-FT establishing itself as the most effective approach for achieving raw accuracy in meter tracking of Carnatic music. Meanwhile, TCN-FS excels in maintaining temporal continuity, particularly in the downbeat tracking task.

The difference in performance between the two models is as expected, given their respective architectures (see Chapter \ref{chap:architectures}) and the application of post-processing in the TCN model. The Beat This! architecture emphasizes the accuracy of local predictions, while the TCN model paired with the postprocessor promotes globally coherent predictions. Additionally, the powerful transformer architecture used by Beat This! is able to extract more meaningful features from the Carnatic data compared to the relatively lightweight TCN model, although this advantage comes with increased computational demands.


\section{Tāḷa-wise Performance}\label{sec:taala_performance}

With TCN-FS and BeatThis-FT identified as the two leading strategies for tracking Carnatic meter, we proceed to analyze their performance on each tāḷa to gain a thorough understanding of their capabilities. Tables \ref{tab:taala_wise_tcnfs} and \ref{tab:taala_wise_btft} provide breakdowns of the performance of TCN-FS and BeatThis-FT, respectively, across the four tāḷas.

\begin{table}[h]
\centering
{\small
\resizebox{\textwidth}{!}{%
\begin{tabular}{lccc|ccc}
\toprule
\textbf{Tāḷa} 
  & \multicolumn{3}{c|}{\textbf{Beat}} 
  & \multicolumn{3}{c}{\textbf{Downbeat}} \\
\cmidrule(lr){2-4} \cmidrule(lr){5-7}
  & \textbf{F-measure} & \textbf{{CML$_t$}} & \textbf{{AML$_t$}} 
  & \textbf{F-measure} & \textbf{{CML$_t$}} & \textbf{{AML$_t$}} \\
\midrule
Ādi (8)            & 77.8 & 52.7 & 84.8 & 62.7 & 42.5 & 84.3 \\
Rūpaka (3)         & 75.8 & 32.8 & 85.0 & 40.7 & 19.5 & 23.8 \\
Mīśra chāpu (7)    & 95.6 & 92.3 & 93.9 & 86.7 & 88.8 & 94.5 \\
Khaṇḍa chāpu (5)   & 93.5 & 84.5 & 88.7 & 68.5 & 64.6 & 65.9 \\
\textbf{Overall}   & 84.6 & 62.9 & 88.0 & 63.9 & 52.1 & 67.0 \\
\bottomrule
\end{tabular}
} 
} 
\caption{TCN-FS : Tāḷa-wise Performance Comparison }
\label{tab:taala_wise_tcnfs}
\end{table}

\begin{table}[h]
\centering
{\small
\resizebox{\textwidth}{!}{%
\begin{tabular}{lccc|ccc}
\toprule
\textbf{Tāḷa} 
  & \multicolumn{3}{c|}{\textbf{Beat}} 
  & \multicolumn{3}{c}{\textbf{Downbeat}} \\
\cmidrule(lr){2-4} \cmidrule(lr){5-7}
  & \textbf{F-measure} & \textbf{{CML$_t$}} & \textbf{{AML$_t$}} 
  & \textbf{F-measure} & \textbf{{CML$_t$}} & \textbf{{AML$_t$}} \\
\midrule
Ādi (8)            & 86.6 & 74.3 & 77.5 & 49.3 & 2.2  & 55.8 \\
Rūpaka (3)         & 89.5 & 74.7 & 77.5 & 81.6 & 68.3 & 68.3 \\
Mīśra chāpu (7)    & 94.2 & 86.0 & 86.1 & 72.8 & 49.7 & 50.6 \\
Khaṇḍa chāpu (5)   & 91.4 & 76.6 & 78.5 & 61.1 & 28.8 & 29.3 \\
\textbf{Overall}   & 90.3 & 78.0 & 80.0 & 66.8 & 38.2 & 53.7 \\
\bottomrule
\end{tabular}
} 
} 
\caption{Beat This-FT : Tāḷa-wise Performance Comparison }
\label{tab:taala_wise_btft}
\end{table}

In terms of beat tracking, BeatThis-FT demonstrates relatively consistent performance across tālas compared to TCN-FS. This consistency is also evident in the continuity scores. TCN-FS struggles particularly with ādi (8) and rūpaka (3) tālas, especially the latter. Although the beat F-measures are reasonable, the low {CML$_t$} scores indicate that while many beats are detected correctly, the system often loses correct tempo continuity throughout the sequence. The higher {AML$_t$} scores suggest that the system frequently predicts tempos that are rhythmically related, implying metrical ambiguity, likely due to the post-processing stage, which is absent in Beat This!. 

Interestingly, even in Beat This!, the tālas ādi (8) and rūpaka (3) score lower in beat tracking accuracy than mīśra chāpu (7) and khaṇḍa chāpu (5). This may seem counter-intuitive, as one might expect systems to struggle more with rarer and more complex meters. However, the difference is due to the variety of patterns within a given tāla. In Carnatic music, performers often improvise and vary grouping structures within a cycle, while maintaining the core framework and overall length.

\begin{figure}[htbp]
    \centering
    \includegraphics[width=\textwidth]{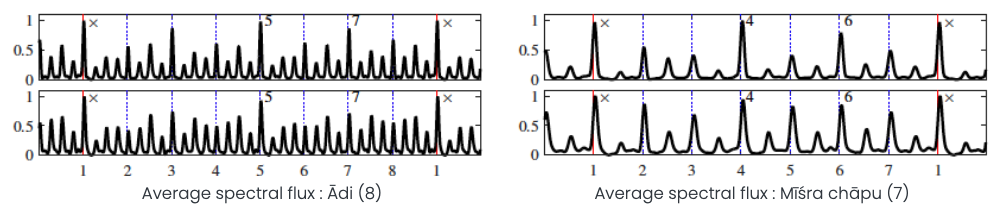}
    \caption[Spectral flux pattern comparison of tālas]{Spectral flux pattern comparison of tālas. Figure taken from \cite{srinivasamurthy_data-driven_2016}.}
    \label{fig:spectral_flux}
\end{figure}

As explained by \cite{srinivasamurthy_data-driven_2016}, multiple rhythmic patterns that depart from the traditional tāla structure can be played. For example, a musician might perform a pattern grouped as 7, 7, 4, 6, and 8 akṣaras, totaling 32 akṣaras within an ādi tāḷa cycle. Popular tālas like ādi and rūpaka tend to have more such variations, making them more difficult for beat tracking systems to generalize. This complexity is visible in the plots in figure \ref{fig:spectral_flux}, which shows the average cycle length spectral flux patterns for ādi and mīśra chāpu tāḷas in the {CMR$_f$} dataset. The patterns indicate varying accent strengths at different metrical positions, reflecting the rhythmic variation within each tāla. Also, both models achieve high performance on mīśra chāpu for both beat and downbeat detection. This consistency implies that mīśra chāpu’s rhythmic pattern is relatively easier to model accurately for both architectures.

When examining the downbeat detection task, the results are more nuanced. While Beat This-FT slightly outperforms TCN-FS in overall F-measure (66.8 vs 63.9), both exhibit mixed results with wide variation across tāḷas. For example, Beat This-FT excels dramatically on rūpaka, whereas it struggles on ādi downbeats. This suggests that each model may be more adept at handling certain rhythmic structures but less consistent across all tāḷa types. For further granularity, Appendix \ref{app:analysis_plots} includes comprehensive violin plots illustrating per-track performance for both training strategies on each tāḷa.


\section{Outlier Analysis}\label{sec:outlier_analysis}

Due to the relatively poor performance of TCN-FS on ādi and rūpaka tāḷas, we perform a preliminary outlier analysis. Tables \ref{tab:worst_tcn} lists the tracks with the lowest beat and downbeat F-measure scores for these tāḷas.

\begin{table}[h]
\centering
{\small
\resizebox{0.65\textwidth}{!}{%
\begin{tabular}{l l r r r}
\toprule
\textbf{track id} & \textbf{tāḷa} 
  & \multicolumn{3}{c}{\textbf{Beat}} \\
\cmidrule(lr){3-5}
  &  & \textbf{F-measure} & \textbf{CML$_t$} & \textbf{AML$_t$} \\
\midrule
10047 & adi & 0.192975 & 0.100241 & 0.458864 \\
11024 & rupakam & 0.605962 & 0.000145 & 0.903054 \\
\midrule
\textbf{track id} & \textbf{tāḷa} 
  & \multicolumn{3}{c}{\textbf{Downbeat}} \\
\cmidrule(lr){3-5}
  &  & \textbf{F-measure} & \textbf{CML$_t$} & \textbf{AML$_t$} \\
\midrule
10048 & adi & 0.000000 & 0.0 & 1.000000 \\
11040 & rupakam & 0.032501 & 0.0 & 0.000000 \\
\bottomrule
\end{tabular}
} 
} 
\caption{TCN-FS : Worst Performing Tracks by Beat and Downbeat F-Measure}
\label{tab:worst_tcn}
\end{table}

\begin{figure}[htbp]
    \centering
    \hspace*{-0.1\textwidth}\includegraphics[width=1.2\textwidth]{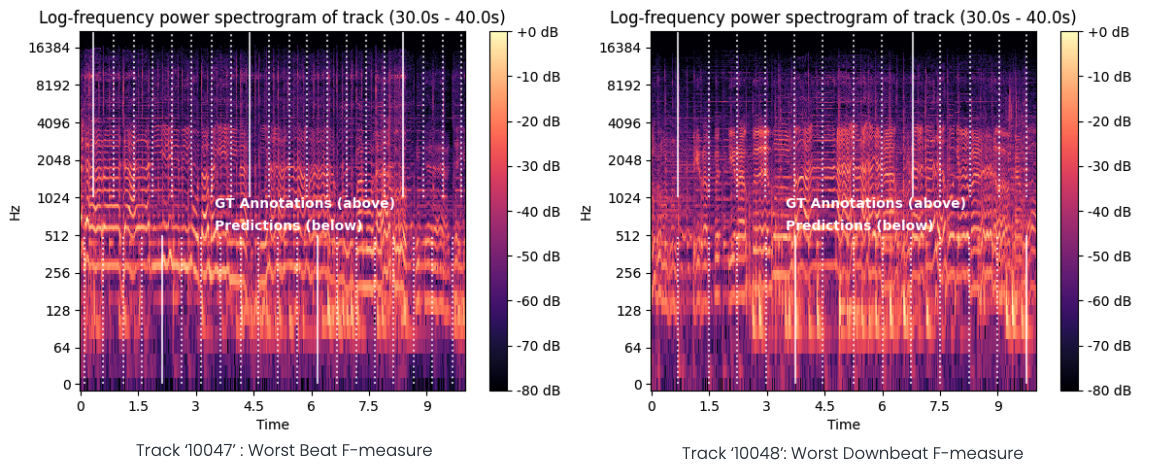}
    \caption[TCN-FS : Worst performing tracks for ādi tāḷa visualised]{TCN-FS : Worst performing tracks for ādi tāḷa visualised}
    \label{fig:tcn_worst}
\end{figure}

Figure \ref{fig:tcn_worst} visualizes the ground truth annotations alongside the model predictions over the spectrogram for sections of the two ādi tāḷa outliers with the lowest beat (track id: 10047) and downbeat F-measure (track id: 10048) scores, respectively. In the first case, it is clear that the beat predictions are consistently shifted by half a beat, resulting in a low beat F-measure score. This is a common occurrence in Carnatic music, where creative phase offsets (\textit{eḍupu}) are often employed - percussive onsets are shifted from the actual beats of the tāḷa (i.e., played on the off-beat) - making it challenging for the network to detect the true beats. Here, the {AML$_t$} score provides a more realistic measure of beat tracking performance.

For the downbeat fail case, the detected downbeats are displaced by exactly half a cycle. Though it scores zero in accuracy, it achieves a perfect {AML$_t$} score. This issue can be attributed to the post-processor, which enforces global prediction constraints and may produce scores unrepresentative of the model’s actual performance. 

The rūpaka outliers were analyzed by listening, as it is more difficult to visually identify the reasons for their poor performance. In the beat tracking outlier, the percussion switches creatively between triple and quadruple meter through metric modulation, challenging the post-processor’s ability to handle these rapid shifts. The downbeat outlier is particularly challenging because the percussion is performed at double tempo compared to the ground truth annotations, while also employing polymeter. The subpar performance of TCN-FS on rūpaka is likely due to these factors and the difficulties they pose for the post-processing stage. 

\section{Tempo and Tāḷa Cycle Duration Effects}\label{sec:tempo_effects}

Lastly, we delve deeper to identify the possible effects of track tempo and tāḷa-cycle duration on model performance. 

\begin{figure}[htbp]
    \centering
    \includegraphics[width=0.8\textwidth]{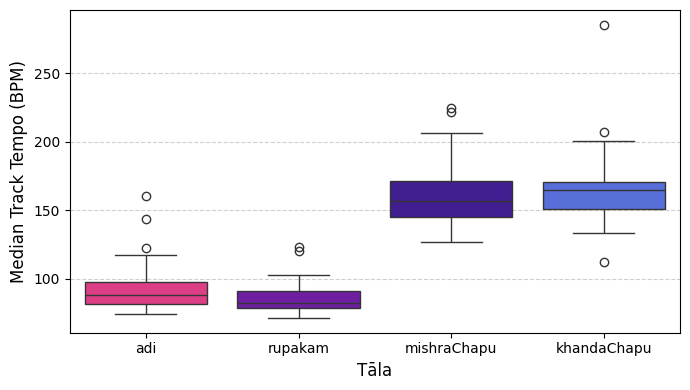}
    \caption[Distribution of median track tempo by tāḷa]{Distribution of median track tempo by tāḷa}
    \label{fig:taala_tempo}
\end{figure}

Figure \ref{fig:taala_tempo} presents a box plot of the median track tempos in the {CMR$_f$} dataset, grouped by tāḷa. For each track, inter-beat intervals (IBI) are converted to BPM values, and the median BPM is plotted. Notably, the majority of tracks in the ādi (8) and rūpaka (3) tāḷas fall within a narrow tempo range of approximately 50 to 100 bpm. In contrast, the two best-performing tāḷas, mīśra chāpu (7) and khaṇḍa chāpu (5), cluster around 160 bpm but exhibit a wider tempo distribution. This stark difference raises an important question: do the ground truth annotations reflect the actual tempo, or are the ādi (8) and rūpaka (3) tracks annotated at half tempo, potentially contributing to their underperformance?

Next, we measure the cycle duration for each track as the interval between consecutive downbeats. Table \ref{tab:cycle_lengths} summarizes the cycle durations by tāḷa. The ādi tāḷa distinctly features longer and slower cycles compared to the others, with a median cycle duration of 5.4 seconds, more than double that of rūpaka (2.1s) and khaṇḍa chāpu (1.8s). This has important implications for downbeat tracking: longer cycle lengths demand a larger context window for accurate detection, which can increase the complexity of the model’s task. Additionally, longer cycles mean fewer downbeat annotations per track, potentially limiting the amount of training data available to the network for learning. Another insight is that the variability in cycle duration is also greatest in ādi tāḷa, with a range from 2.9 to 7.1 seconds.

\begin{table}[h]
\centering
\resizebox{0.75\textwidth}{!}{%
\begin{tabular}{lccc}
\toprule
\textbf{Tāla} & \textbf{Min. cycle} & \textbf{Max. cycle} & \textbf{Median cycle} \\
             & \textbf{duration (s)} & \textbf{duration (s)}  & \textbf{duration (s)} \\
\midrule
Ādi (8)           & 2.9  & 7.1  & 5.4 \\
Rūpaka (3)        & 1.2  & 3.1  & 2.1 \\
Miśra Chāpu (7)   & 1.6  & 3.6  & 2.6 \\
Khaṇḍa Chāpu (5)  & 0.9  & 2.9  & 1.8 \\
\bottomrule
\end{tabular}%
}
\caption{Tāla-wise summary of cycle durations (in seconds).}
\label{tab:cycle_lengths}
\end{table}

To investigate the potential effects of tempo and cycle duration on model performance, we plotted track tempo and cycle duration against beat and downbeat accuracy for both TCN-FS and BeatThis-FT. These plots, provided in Appendix \ref{app:analysis_plots}, also reflect variability in tempo and cycle duration through the point sizes. Overall, no conclusive evidence emerged linking model performance directly with tempo or cycle duration variations.

Nonetheless, TCN-FS, which relies on post-processing, can benefit from informed tempo constraints that narrow the search space and reduce ambiguities like tempo octave errors. Currently, the post-processor operates in a broad range of tempos to accommodate the four tāḷas. However, narrowing this range based on tempo analysis of the training data, especially in tāḷa-informed meter tracking, is likely to enhance performance. This highlights the importance of tempo profiling as a preparatory step for post-processor dependent tracking systems.

\chapter{Conclusions and Future Work}\label{chap:conclusions_future}

\section{Summary of the Study}\label{sec:study_summary}

In this study, we investigated the applicability and adaptation of state-of-the-art deep learning models for meter tracking in Carnatic music, a rich and rhythmically complex tradition from the Indian subcontinent. 

Our methodology involved establishing a baseline using traditional Dynamic Bayesian Networks (DBNs) to set a performance benchmark on the Carnatic Music Rhythm dataset ({CMR$_f$}). We evaluated two modern neural architectures: Temporal Convolutional Network (TCN) that utilises post-processing with a DBN to enhance temporal coherence, and a transformer-based model called Beat This!, which achieves meter tracking without relying on any post-processing. For each, we tested various training strategies, including training from scratch on Carnatic data and fine-tuning models pre-trained on Western music datasets. We incorporated musically informed techniques aimed at improving performance and adaptation, such as tāḷa-aware training splits and post-processing adaptations, where applicable.

The two models exhibited complementary strengths: Beat This! fine-tuned on Carnatic music excelled in raw beat and downbeat accuracy without post-processing, while, the TCN model trained from scratch on Carnatic music combined with post-processing better maintained temporal continuity, notably in downbeat tracking. The choice between the two may also depend on the specific tāḷa and whether the priority is accuracy or continuity of detections, such as in real-time applications.

Analysis across different tāḷas revealed the strengths and weaknesses of the two models and highlighted challenges: popular tāḷas such as ādi and rūpaka, characterized by diverse rhythmic variations, posed greater difficulties compared to mīśra chāpu and khaṇḍa chāpu. Further analysis revealed that rhythmic complexities, including creative phase offsets and polymetric structures, frequently led to metrical ambiguities, especially in post-processor-based systems, which in turn diminished their evaluation metrics.

\section{Conclusions}\label{sec:conclusions}

The results of this study demonstrate that modern deep learning models tailored or fine-tuned specifically for Carnatic music considerably outperform the traditional DBN baseline in both beat and downbeat tracking. Models trained only on Western repertoires lack sufficient generalizability to handle the diverse and intricate rhythmic patterns found in Carnatic music. This highlights the importance of domain-specific adaptation, where even modest amounts of annotated data can substantially enhance model performance through transfer learning.

Transformer-based models such as Beat This! excel at capturing global rhythmic structure but are data- and resource-intensive, whereas TCNs combined with post-processing are more lightweight and practical for real-time or resource-constrained applications. Designers of meter tracking systems must therefore consider the balance between predictive accuracy, temporal coherence, and computational efficiency when selecting models.

A key observation is the trade-off between reliance on post-processing and the intrinsic predictive ability of models. While music-informed post-processing is advantageous, it creates dependencies that mask the network’s intrinsic performance. This strengthens the case for post-processor-free models and evaluation methods that more accurately reflect raw network capabilities, enabling a clearer assessment of deep learning architectures independent of hand-crafted constraints.

Overall, this study shows that tailored training strategies and model designs are essential for capturing the complex rhythms of Carnatic music. More broadly, it highlights how deep learning models can be adapted to underrepresented non-Western traditions when musical context, data, and application needs are carefully considered, advancing automatic rhythm analysis in diverse cultural settings.

\section{Future Work}\label{sec:future_work}

Building on the insights and limitations identified, future research directions include:

\begin{itemize}
    \item \textbf{Model Tuning and Ablation:} Systematic experimentation with existing model hyperparameters, such as varying dilation rates and the number of filters in TCN, alongside conducting ablation studies, can identify parts of the model that matter most for better performance.
    \item \textbf{Hybrid Modeling Strategies:} Exploring hybrid strategies that integrate the strengths of both TCN and Beat This! may yield systems that balance sharp prediction accuracy with temporal coherence. For instance, utilising \textit{shift-tolerant loss} and \textit{sum heads} in TCN to encourage strong local predictions reducing the reliance on post-processing. 
    \item \textbf{Data Augmentation and Synthetic Expansion:} Targeted augmentation techniques could enrich the training datasets, particularly benefiting tāḷas with greater pattern variations and challenging rhythmic structures.
    \item \textbf{Advanced Evaluation Metrics:} Developing or adopting metrics better suited to capture expressive timing and metrical nuances in Carnatic music will offer more meaningful performance assessments.
    \item \textbf{Cross-Tradition Generalization:} Extending research to North Indian Hindustani music and other underrepresented non-Western music traditions could validate the generalizability of adaptation strategies and enrich ethnomusicological MIR research.
\end{itemize}

Collectively, these future directions aim to enhance the inclusiveness, accuracy, and practical adoption of deep learning-based automatic meter tracking systems tailored to the rich rhythmic diversity present in Carnatic music and beyond.

\newpage
\listoffigures
\listoftables

\bibliographystyle{plainnat}
\bibliography{preamble/bibliography}

\appendixpageoff
\begin{appendices}
\chapter{Software and Other Resources}\label{app:supplementary}

This appendix provides a comprehensive list of key software resources, datasets, and code repositories utilised in this study. These materials enable reproducibility of the experiments and analyses presented. Additionally, relevant reference works and helpful study resources are included for further exploration.

\textbf{Datasets}

The \textit{Carnatic Music Rhythm} (CMR$_f$) dataset \cite{srinivasamurthy2014supervised} is available for download upon request from the CompMusic Project website: \\
\href{https://compmusic.upf.edu/carnatic-rhythm-dataset}{\textcolor{blue}{https://compmusic.upf.edu/carnatic-rhythm-dataset}}

Links to important Western music datasets commonly used for meter tracking, some of which were utilised for pretraining models in this study, can be accessed at: \\
\href{https://ismir.net/resources/datasets/}{\textcolor{blue}{https://ismir.net/resources/datasets/}}

\vspace{0.3cm}

\textbf{Reproducible Code}

The codebase for training the Temporal Convolutional Network (TCN) on the CMR$_f$ dataset is available at: \\
\href{https://github.com/satyajeetprabhu/tcn-carnatic-tracker}{\textcolor{blue}{https://github.com/satyajeetprabhu/tcn-carnatic-tracker}}

The repository for fine-tuning the \textit{Beat This!} model on the CMR$_f$ dataset can be found at: \\
\href{https://github.com/satyajeetprabhu/beat-this-carnatic}{\textcolor{blue}{https://github.com/satyajeetprabhu/beat-this-carnatic}}

Both repositories include the trained models from this study, evaluation results, and notebooks for reproducing the analyses and plots.

\vspace{0.3cm}

\textbf{Reference Implementations}

The TCN implementation employed, developed in PyTorch Lightning, is based on the \textit{LAMIR 2024 Hackathon Tutorial} \cite{morais_lamir_2024} on adapting deep learning models for Latin American music tasks with limited data: \\
\href{https://lamir-workshop.github.io/lamir_hackathon/intro.html}{\textcolor{blue}{https://lamir-workshop.github.io/lamir\_hackathon/intro.html}}

The original \textit{Beat This!} \cite{foscarin_beat_2024} implementation is available at: \\
\href{https://github.com/CPJKU/beat_this}{\textcolor{blue}{https://github.com/CPJKU/beat\_this}}

The fine-tuning code for \textit{Beat This!} was adapted from work by SMC Master students Milo Beuzeval and Navid Hallajian: \\
\href{https://github.com/smilo7/more-beats-for-this}{\textcolor{blue}{https://github.com/smilo7/more-beats-for-this}}

\vspace{0.3cm}

\textbf{Key Software Libraries}

The \textit{madmom} Python audio and music signal processing library \cite{madmom} used for audio preprocessing tasks: \\
\href{https://github.com/CPJKU/madmom}{\textcolor{blue}{https://github.com/CPJKU/madmom}}

The \textit{madmom} Dynamic Bayesian Network (DBN) post-processor \cite{bock_joint_2016, krebs_efficient_2015} employed alongside the TCN: \\
\href{https://madmom.readthedocs.io/en/v0.16/modules/features/downbeats.html}{\textcolor{blue}{https://madmom.readthedocs.io/en/v0.16/modules/features/downbeats.html}}

The \textit{mirdata} Python library \cite{bittner_fuentes_2019} used for dataset loading, validation, and parsing: \\
\href{https://github.com/mir-dataset-loaders/mirdata}{\textcolor{blue}{https://github.com/mir-dataset-loaders/mirdata}} \\
Documentation: \\
\href{https://mirdata.readthedocs.io/en/stable/}{\textcolor{blue}{https://mirdata.readthedocs.io/en/stable/}}

The \textit{mir\_eval} Python library \cite{Raffel2014mir_eval} used for evaluation: \\
\href{https://github.com/mir-evaluation/mir_eval}{\textcolor{blue}{https://github.com/mir-evaluation/mir\_eval}} \\
Documentation: \\
\href{https://mir-eval.readthedocs.io/latest/}{\textcolor{blue}{https://mir-eval.readthedocs.io/latest/}}

\vspace{0.3cm}

\textbf{Additional Study Resources}

The ISMIR 2021 tutorial on tempo, beat, and downbeat estimation \cite{tempobeatdownbeat:book} provides a comprehensive overview of deep learning models for beat and downbeat tracking. It also includes an open-source, TensorFlow-based implementation of the TCN model described in \textit{Deconstruct, Analyse, Reconstruct} \cite{bock_deconstruct_2020}, which is the basis for the TCN model employed in this study: \\
\href{https://tempobeatdownbeat.github.io/tutorial/intro.html}{\textcolor{blue}{https://tempobeatdownbeat.github.io/tutorial/intro.html}}

The Python notebooks accompanying the textbook \textit{Fundamentals of Music Processing (FMP)} \cite{muller_fundamentals_2021} provide foundational material on computational music analysis using signal processing techniques. Chapter 6 (Tempo and Beat Tracking) is of special relevance to this study: \\
\href{https://www.audiolabs-erlangen.de/resources/MIR/FMP/C6/C6.html}{\textcolor{blue}{https://www.audiolabs-erlangen.de/resources/MIR/FMP/C6/C6.html}}

\chapter{Detailed Analysis Plots}\label{app:analysis_plots}
\begin{figure}[htbp]
    \centering
    \hspace*{-0.05\textwidth}\includegraphics[width=1.1\textwidth]{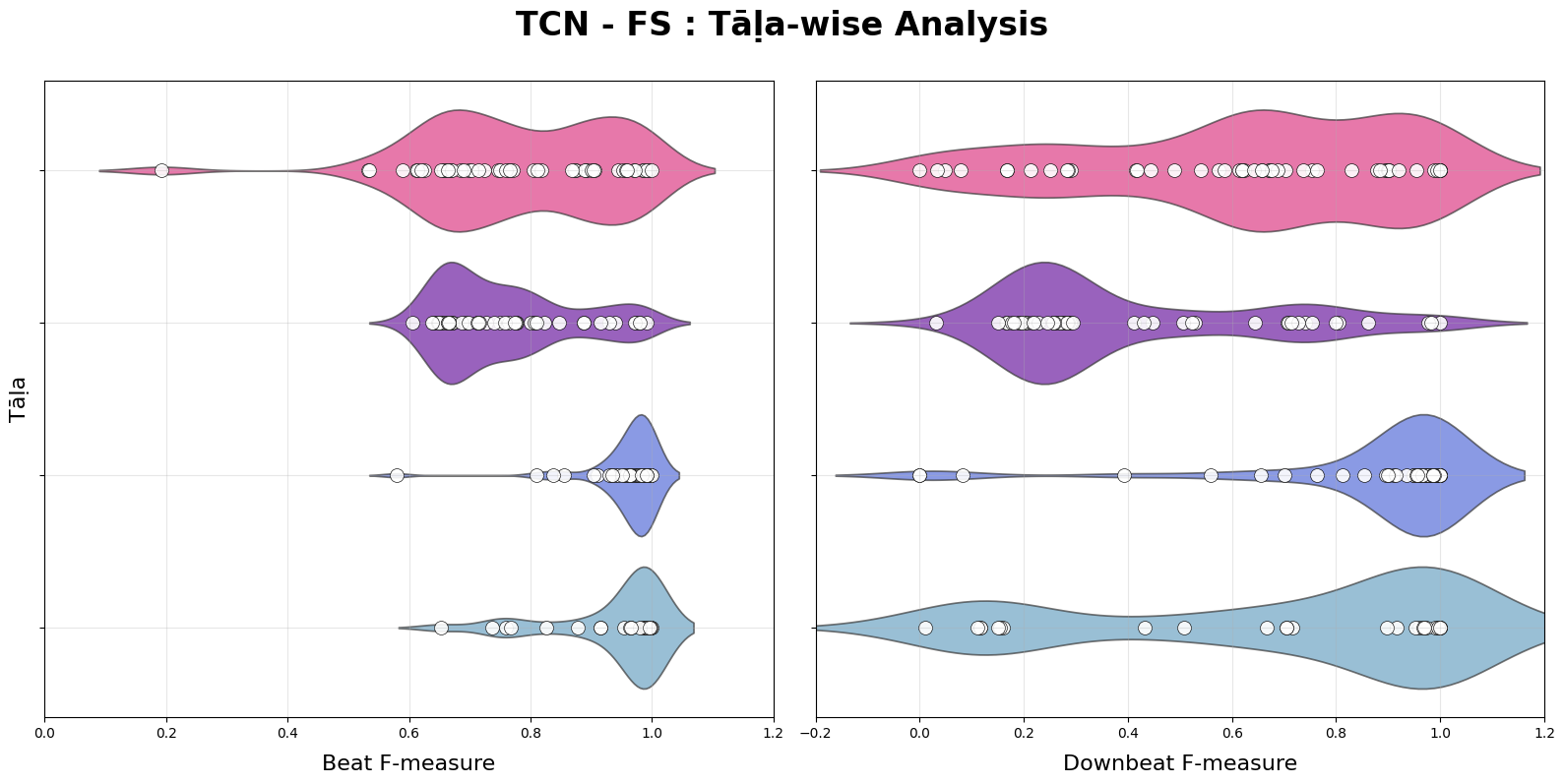}
    \hspace*{-0.05\textwidth}\includegraphics[width=1.1\textwidth]{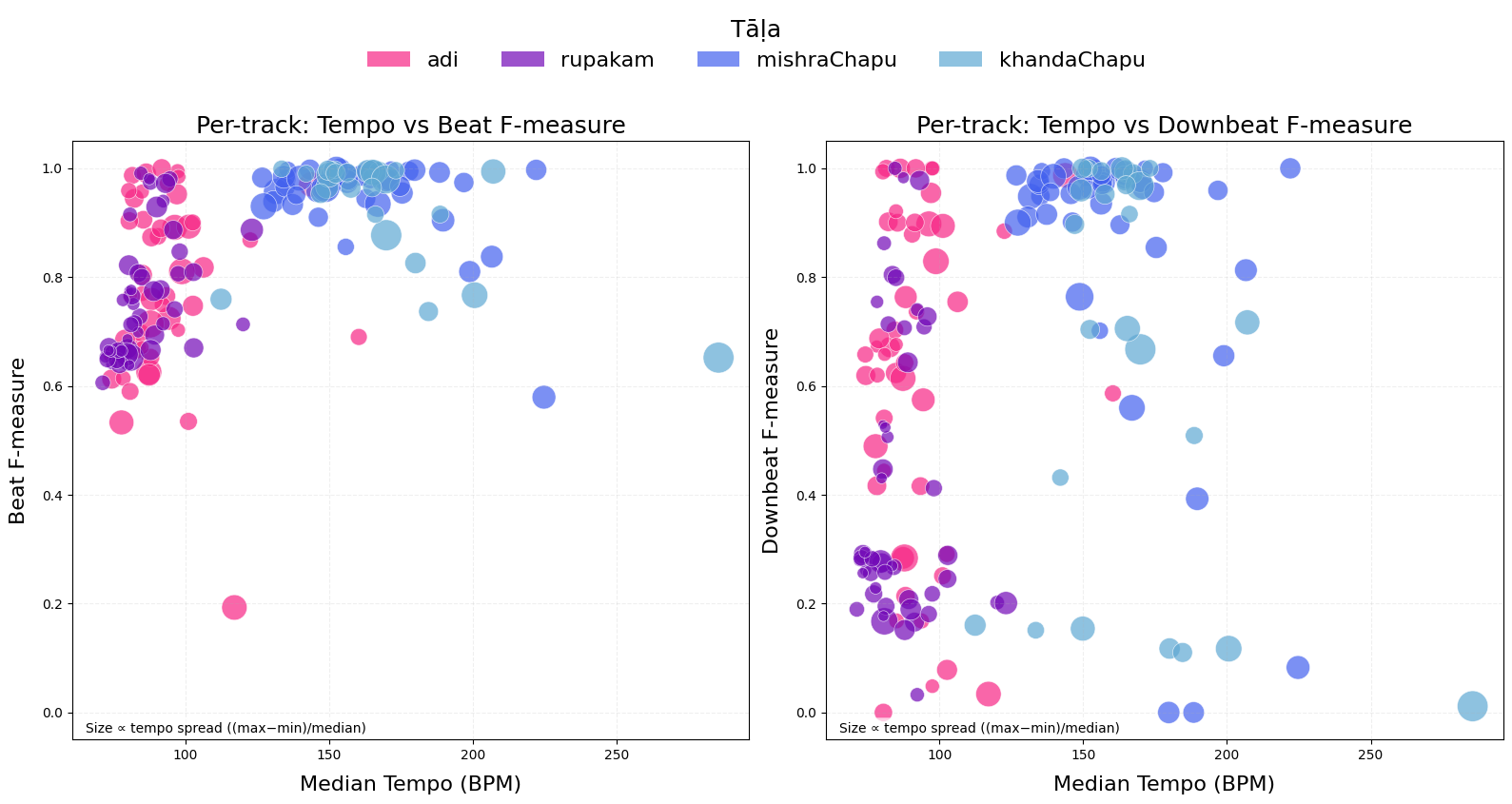}
    \hspace*{-0.05\textwidth}\includegraphics[width=1.1\textwidth]{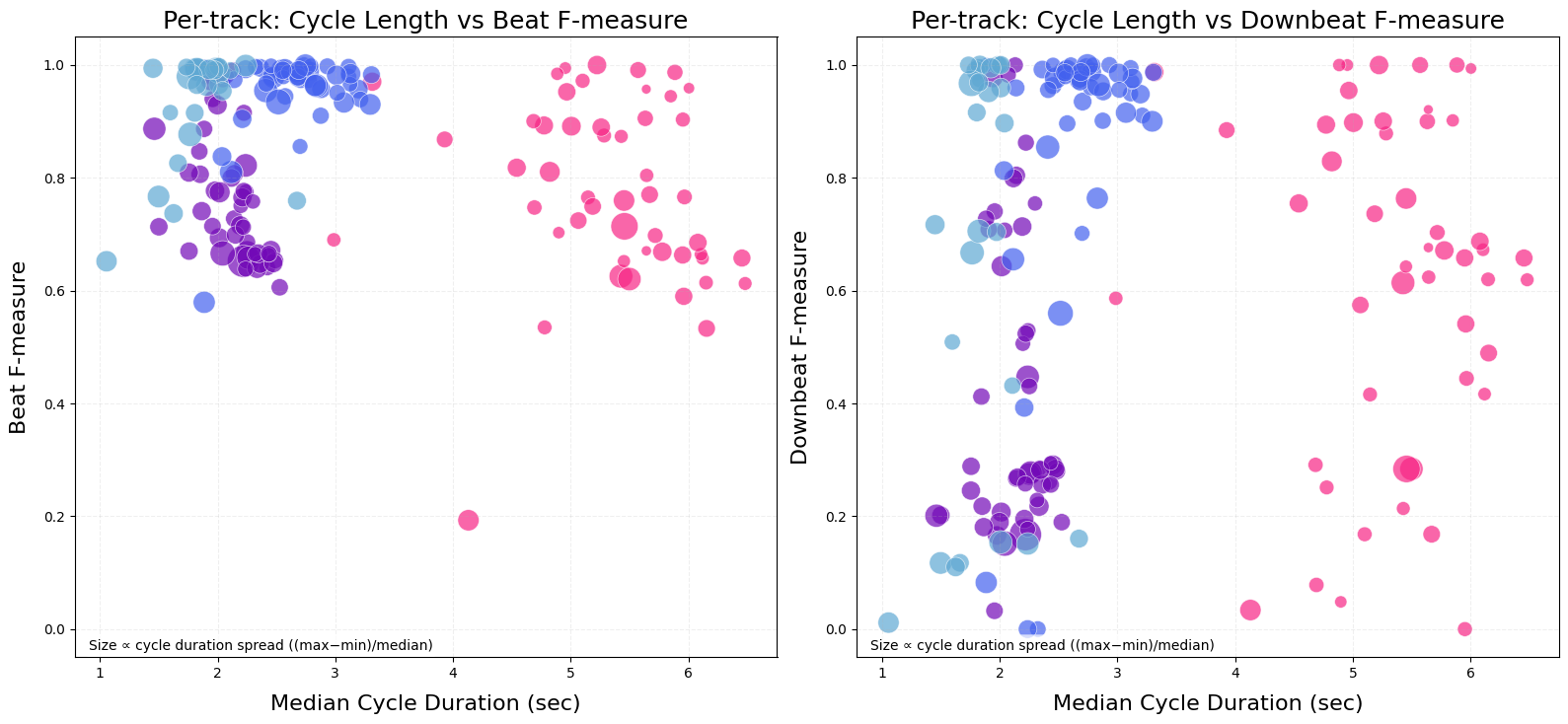}
\end{figure}

\begin{figure}[htbp]
    \centering
    \hspace*{-0.05\textwidth}\includegraphics[width=1.1\textwidth]{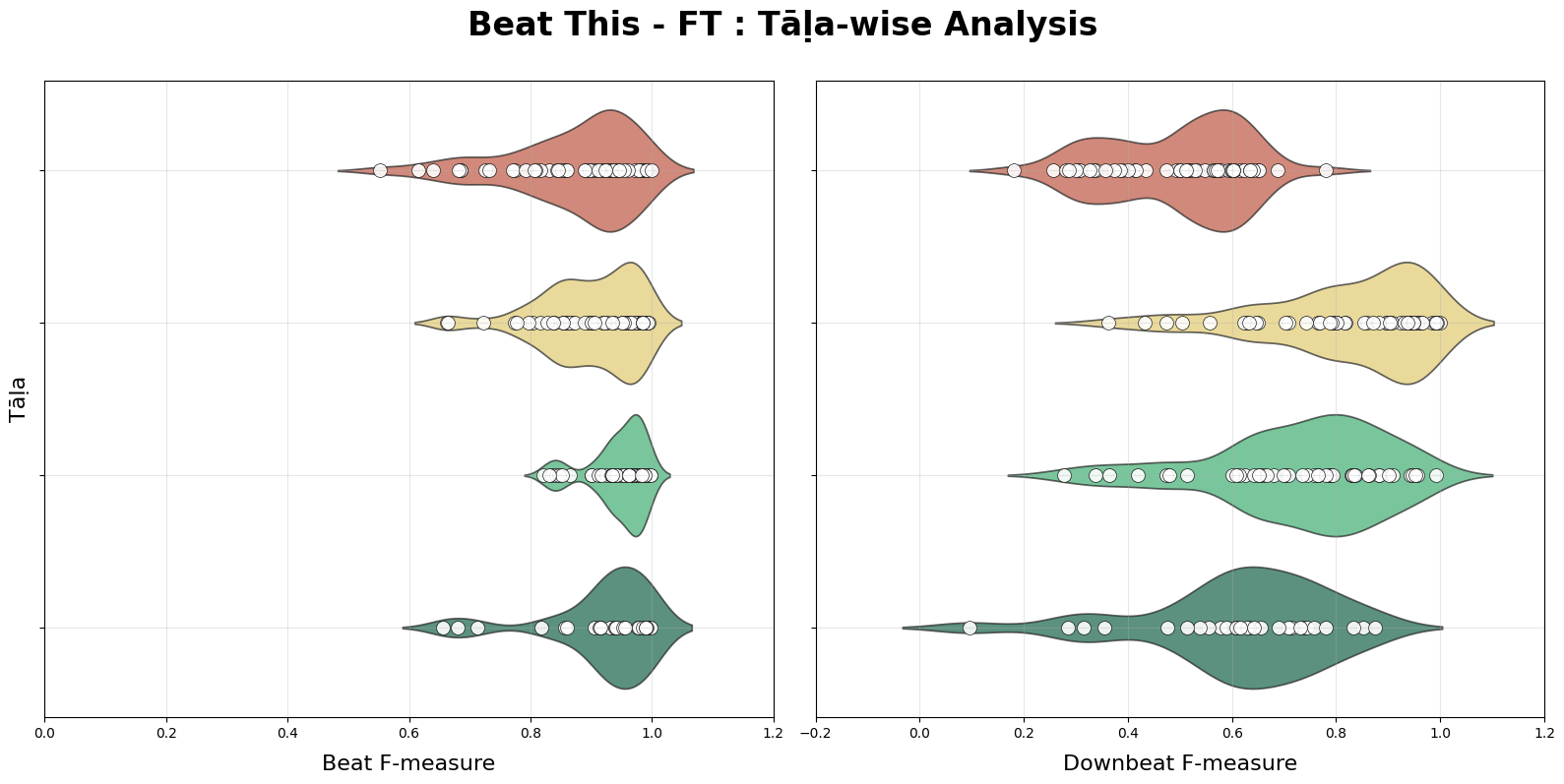}
    \hspace*{-0.05\textwidth}\includegraphics[width=1.1\textwidth]{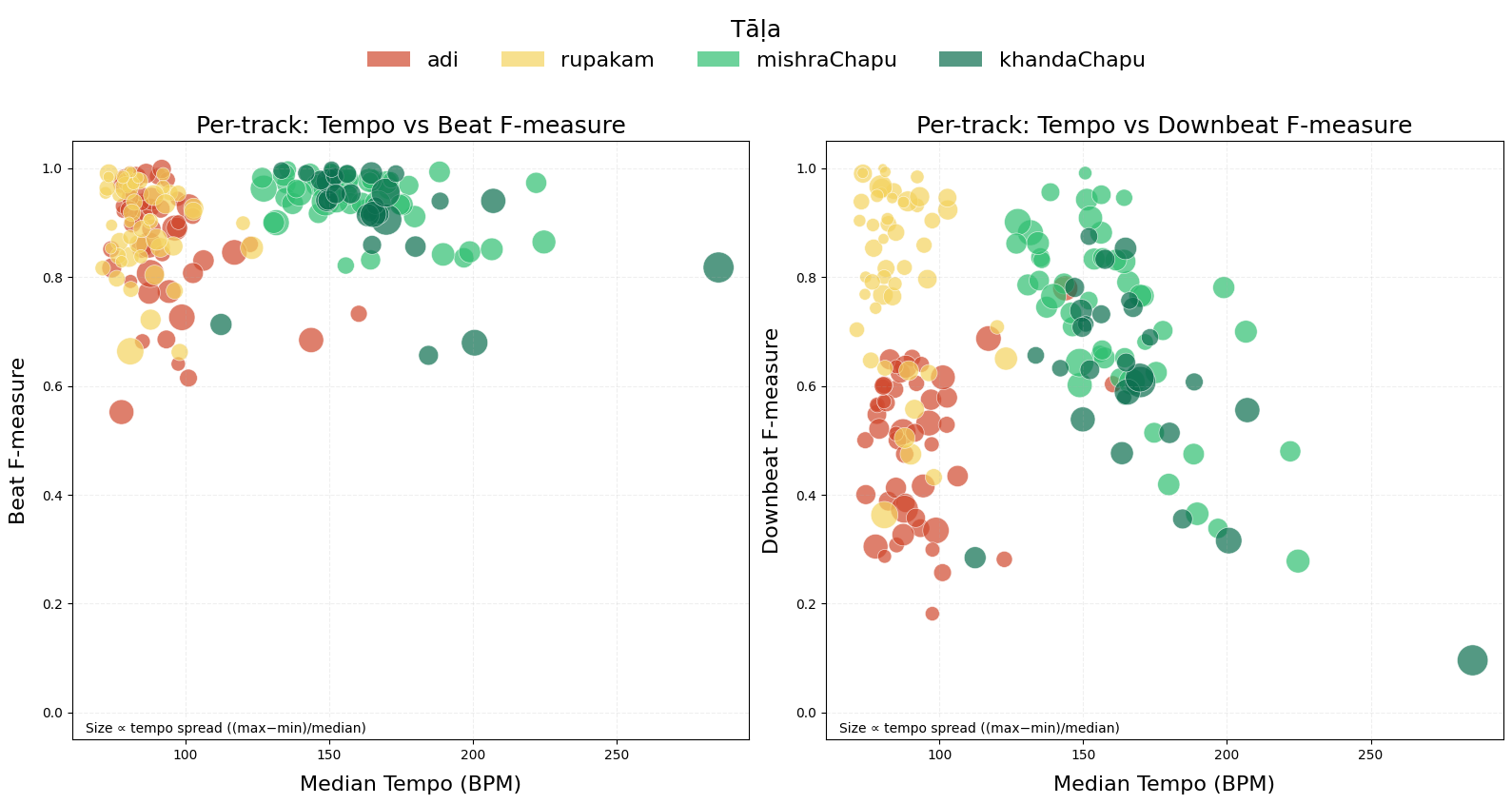}
    \hspace*{-0.05\textwidth}\includegraphics[width=1.1\textwidth]{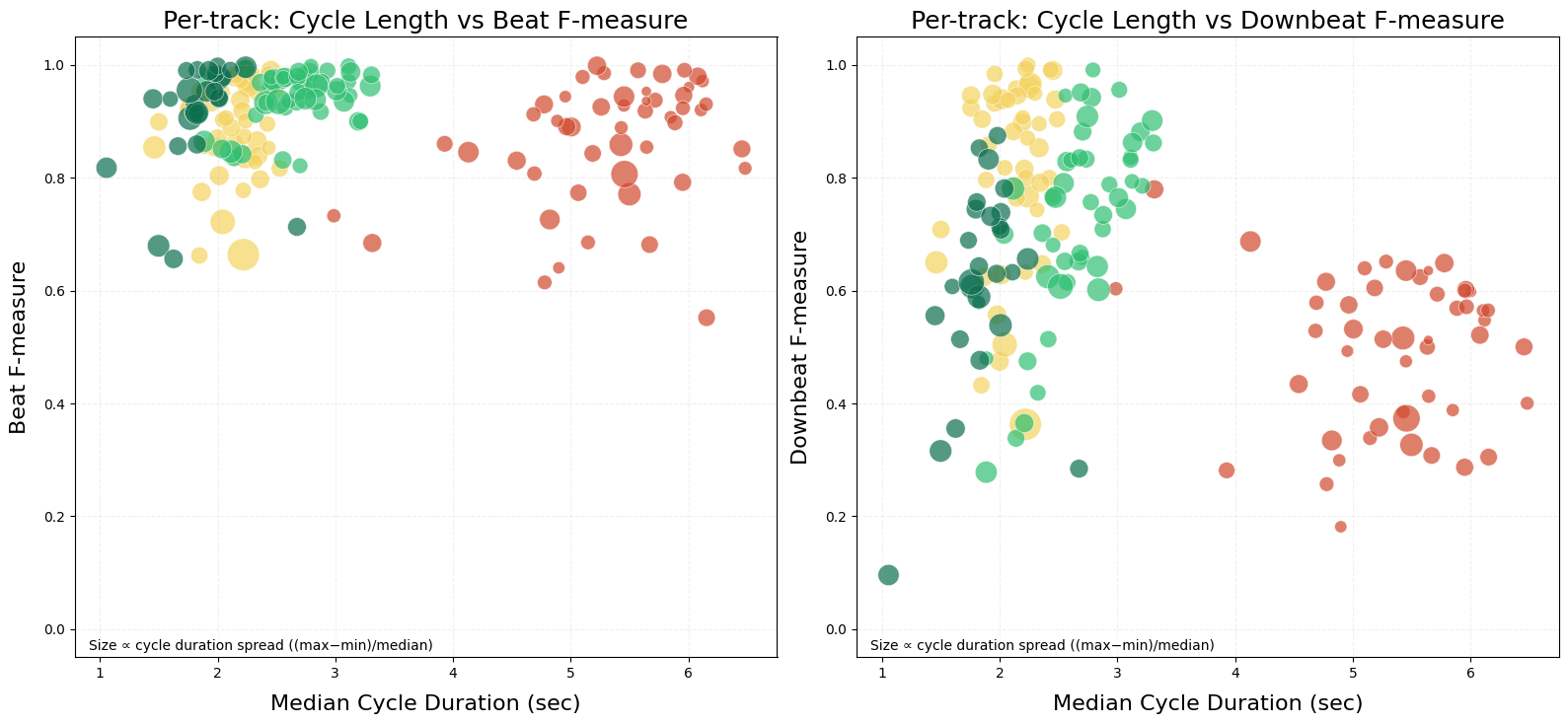}
\end{figure}

\end{appendices}

\end{document}